\documentclass[11pt,a4paper]{article}
\usepackage{epsfig}
\usepackage{graphics}
\usepackage{graphicx}
\usepackage{pstricks,pst-coil,pst-fill,pst-plot}
\usepackage{cite,indentfirst}
\usepackage{amsmath}    
\usepackage{amssymb}    
\usepackage{amsfonts}   
\usepackage{verbatim}   
\usepackage{mathrsfs}   
\usepackage{dsfont}
\usepackage{vmargin}        
\usepackage{euscript}   

\setmarginsrb{3.4cm}{2.5cm}{3.4cm}{2.5cm}{1cm}{1cm}{1cm}{1.6cm}

\def\C|{{\mathbb C} \,}
\def\B|{{\mathbb B} \,}
\def\S|{{\mathbb S} \,}
\def\G|{{\mathbb G} \,}
\def\N|{{\mathbb N} \,}
\def\F|{{\mathbb F} \,}
\def\K|{{\mathbb K} \,}

\def\tens{\mathop\otimes\limits}

\def\sg{\sigma}

\def\sgz{\sigma^z}

\def\eps{\varepsilon}

\newcommand{\bra}[1]{\langle #1\,|}
\newcommand{\ket}[1]{|#1\,\rangle}

\let\tend=\rightarrow

\newtheorem{theorem}{Theorem}[section]
\newtheorem{prop}{Proposition}[section]
\newtheorem{cor}{Corollary}[section]

  \newtheorem{lemme}{Lemma}
\def\Proof{\medskip\noindent {\it Proof --- \ }}

\let\qed=\cqfd

\renewcommand{\theequation}{\thesection.\arabic{equation}}
\newcommand\beq{\begin{equation}}
\newcommand\enq{\end{equation}}

\def\beqa{\begin{eqnarray}}
\def\eeqa{\end{eqnarray}}
\def\ba{\begin{array}}
\def\ea{\end{array}}

\def\a{\alpha}
\def\b{\beta}

\def\det{\operatorname{det}}
\def\sul{\sum\limits}
\def\pl{\prod\limits}
\def\lt({\left(}
\def\rt){\right)}

\numberwithin{equation}{section}

\def\eps{\epsilon}

\def\la{\lambda}
\def\sg{\sigma}

\newcommand{\f}[2]{{\ensuremath{%
    \mathchoice%
    {\dfrac{#1}{#2}}
    {\dfrac{#1}{#2}}
    {\frac{#1}{#2}}
    {\frac{#1}{#2}}
}}}
\newcommand{\tf}[2]{\ensuremath{#1/#2}}
\newcommand{\pa}[1]{\ensuremath{\left(#1\right)}}
\newcommand{\paa}[1]{\ensuremath{\left\{#1\right\}}}
\newcommand{\pac}[1]{\ensuremath{\left[#1\right]}}
\newcommand{\paf}[2]{\ensuremath{\left(\f{#1}{#2}\right)}}

\newcommand{\mc}[1]{\ensuremath{\mathcal{#1}}}

\newcommand{\ov}[1]{\ensuremath{\overline{#1}}}

\newcommand{\Int}[2]{\ensuremath{\int\limits_{#1}^{#2}}}

\newcommand{\R}{\ensuremath{\mathbb{R}}}
\newcommand{\Cset}{\ensuremath{\mathbb{C}}}

\newcommand{\s}[1]{\ensuremath{\sinh\pa{#1}}}
\newcommand{\sd}[1]{\ensuremath{\mathfrak{s}\pa{#1}}}

\newcommand{\ex}[1]{\ensuremath{\mathrm{e}^{#1}}}

\newcommand{\abs}[1]{\ensuremath{\mid #1 \mid}}

\newcommand{\moy}[1]{\ensuremath{\langle #1 \rangle}}

\renewcommand{\det}[2]{\ensuremath{\mathrm{det}_{#1}\pac{#2}}}

\newcommand{\dd}{\mathrm{d}}
\newcommand{\e}[1]{\ensuremath{\mathrm{#1}}}

\newcommand{\intoo}[2]{\ensuremath{\left ] \, #1 \,; #2 \, \right [ }}

\begin{document}



\begin{flushright}
LPENSL-TH-03/08\\
\end{flushright}
\par \vskip .1in \noindent

\vspace{24pt}

\begin{center}
\begin{LARGE}
{\bf   Correlation functions of the open XXZ chain II}
\end{LARGE}

\vspace{50pt}

\begin{large}

{\bf N.~Kitanine}\footnote[1]{LPTM, Universit\'e de Cergy-Pontoise et CNRS, France,
kitanine@ptm.u-cergy.fr},~~
{\bf K.~K.~Kozlowski}\footnote[2]{ Laboratoire de Physique, Universit\'e de Lyon, ENS Lyon et CNRS,  France,
 karol.kozlowski@ens-lyon.fr},~~
{\bf J.~M.~Maillet}\footnote[3]{ Laboratoire de Physique, Universit\'e de Lyon, ENS Lyon et CNRS,  France,
 maillet@ens-lyon.fr},\\
\vspace{0.1cm}
{\bf G.~Niccoli}\footnote[4]{DESY Theory group, Hamburg, Germany, Giuliano.Niccoli@ens-lyon.fr},~~
{\bf N.~A.~Slavnov}\footnote[5]{ Steklov Mathematical Institute, Moscow, Russia, nslavnov@mi.ras.ru},~~
{\bf V.~Terras}\footnote[6]{ Laboratoire de Physique, Universit\'e de Lyon, ENS Lyon et CNRS,  France, veronique.terras@ens-lyon.fr,
on leave of absence from LPTA, Universit\'e Montpellier II et CNRS, France} \par

\end{large}
\end{center}

\vspace{3cm}

\begin{abstract}
We derive compact multiple integral formulas for several physical spin correlation functions in the
semi-infinite XXZ chain with a longitudinal boundary magnetic field.
Our formulas follow from several  effective re-summations of the multiple integral representation
for the elementary blocks obtained in our previous article (I). In the free fermion point we compute the local magnetization as well as the density of energy profiles. These quantities,
in addition to their bulk behavior, exhibit Friedel type oscillations induced by the boundary; their amplitudes depend on the boundary magnetic field and decay algebraically in terms of the
distance to the boundary.
\end{abstract}

\section{Introduction}

The Hamiltonian of the
Heisenberg XXZ spin-$\tf{1}{2}$ finite chain \cite{Hei28} with diagonal boundary
conditions (namely with longitudinal boundary magnetic fields) is defined as \cite{AlcBBBQ87,Skl88}
\begin{equation}\label{HamXXZ}
 \mc{H} =\sum_{m=1}^{M-1} \paa{  \sigma^x_m \sigma^x_{m+1} +
  \sigma^y_m\sigma^y_{m+1} + \Delta\pa{\sigma^z_m\sigma^z_{m+1}-1} }
  +h_-\sgz_1 +h_+\sgz_M.
\end{equation}
%
%
This is a linear operator acting in the quantum space
$\EuScript{H}=\tens_{m=1}^{M} \EuScript{H}_m , \,\, \EuScript{H}_m \simeq \Cset^2$, of dimension $2^M$ of the chain.
In this expression, $\sg_m^\pm$, $\sg_m^z$ denote local spin operators (acting as Pauli matrices) at site $m$, $\Delta$ is the anisotropy parameter and $h_\pm$ are the boundary (longitudinal) magnetic fields.

We have recently developed a method to compute the so-called
elementary blocks of correlation functions for this model (see \cite{KitKMNST07}, that we refer to as Paper I in the following) in the framework of the (algebraic) Bethe ansatz \cite{Bet31,Hul38,Orb58,Wal59,YanY66,YanY66a,LieM66L,FadST79,FadT79,Tha81,Bax82L, Gau83L, BogIK93L, JimM95L} for boundary integrable
systems \cite{Gau71,AlcBBBQ87,Skl88,Che84,PasS90,KulS91,MezN91,KulS92,GhoZ94,FenS94,SkoS95,LecMSS95,SaT95,KapS96,LesSS96}. The results  essentially agree with previous expressions derived from the vertex operator approach \cite{JimKKKM95, JimKKMW95}.
The purpose of the present paper is to obtain
the physical spin correlation functions for this model, in particular, the one point functions
for the local spin operators at distance $m$ from the boundary as well as several two point functions (like boundary-bulk correlation functions).
There are numerous physical interests in such quantities that can be measured in actual experiments \cite{Fuj03,FujE04,FurH04,HikF04,BorS05,GohBF05,SirB06,SirB06a,SirLFEA07,ChaKBDMEAY04,EggG95,FabG95,BedBFN}.

In much the same way as in the bulk case \cite{KitMT99,KitMT00,KitMST02a,KitMST02b,KitMST04a,KitMST04c}, the computation of the  physical correlation functions amounts to obtain effective re-summations of the multiple integral
representations derived for the elementary blocks. For example,
the one point functions at distance $m$ from the boundary, such as the local
magnetization $\moy{\sg_m^z}$, can be written as a sum of $2^m$ elementary blocks.
We will show how to obtain compact expressions for such objects, typically involving the
sum of only $m$ terms, each containing multiple integrals whose integrants have
a structure similar  to the one of the elementary blocks. In the free fermion point
we are able to compute these multiple integrals (and hence the corresponding
correlations functions) almost completely by reducing them to single integrals. For instance the local magnetization and  the density of energy profiles
(a quantity of interest in the study and the understanding of the interplay
between quantum entanglement and quantum criticality
 \cite{VidLRK03,CalC04,CalC05,OstAFF02,OsbN02,RefM04,Kor04,Laf05,LafSCA06}) are  expressed as single integrals. Hence, their asymptotic behavior at long distance $m$ from the boundary
can be explicitly evaluated. In addition to the bulk constant value they exhibit Friedel
type oscillations \cite{EggG95,FabG95,BedBFN,Laf05,LafSCA06}, algebraically decaying with the distance $m$, their amplitudes being rational functions of the boundary magnetic field,
 in agreement with  field theory predictions \cite{Aff88,AffL91,EggA92,EggA95,WesH00,BilW99,Bil00,Bil00a,WhiAS02,BorS05,GohBF05,SirB06,SirB06a,SirLFEA07}.

We start this paper with a short technical introduction concerning the algebraic Bethe ansatz approach to the open
XXZ spin-$\tf{1}{2}$ chain subject to diagonal boundary magnetic fields.
This preliminary section is followed in Section~\ref{sec-method} by a reminder of the method
proposed in~\cite{KitKMNST07} to compute
correlation functions of open integrable models in the framework of algebraic Bethe ansatz.
In Section~\ref{Action of local operators on boundary states} we
obtain formulae for the action of local operators on arbitrary
boundary states in a form suitable for taking later on the thermodynamic limit. Using these results, we derive a series representation for
the generating function  $\moy{\mc{Q}_m\pa{\kappa}}$ of bulk-boundary $\sg^z$
correlation functions in Section~\ref{Q kappa a limite M infini}. This formula is the boundary analogue of the original
series \cite{KitMST02a} in the bulk case. In Section \ref{Resommation alternative}, we obtain
a formula for
$\moy{\mc{Q}_m\pa{\kappa}}$ alternative to the one inferred in  Section \ref{Q
kappa a limite M infini}. We also  give multiple integral representations  for
$\moy{\sg_{m+1}^+\sg_1^{1}}$ and for the local density of energy.
 These formulae are obtained by a direct resummation of the corresponding elementary blocks. It is worth
stressing that we actually have two representations for their integrand. The first one
is in the spirit of the bulk case \cite{KitMST04c} and involves the Izergin determinant representation \cite{Ize87}
for the partition function  of the six vertex model with domain wall boundary conditions. The second
one involves the Tsuchiya \cite{Tsu98} determinant representation for the partition function of the six vertex model with reflecting ends.
The next section is devoted to the free fermion point. For that case are able to
reduce the multiple integrals to one dimensional ones. This allows us to write the leading asymptotics of
the local magnetization and of the density of energy profiles as well as of the $\moy{\sg_{m+1}^+\sg_1^-}$ correlation function. Our conclusions are presented in the last section.

\section{The open XXZ spin-$\tf{1}{2}$ chain}
\label{technical introduction}

The spectrum of $\mc{H}$ can be obtained by algebraic Bethe
ansatz (ABA) \cite{Skl88}. The central tool
of this method is the boundary monodromy matrix, which will be defined after we introduce some necessary notations.

Here and in the following we adopt the standard parameterizations
$\Delta=\cosh\eta$  and $h_{\pm}= \sinh{\eta} \coth\xi_{\pm}$.

Let $R: \Cset  \tend \mathrm{End}(V\tens V)$, $V\simeq \Cset^2$,
be the  $R$-matrix of the six-vertex model, obtained as the trigonometric solution of the Yang-Baxter equation:
\beq
   R(u)= \s{u+\eta}\,\widehat{R}(u),\qquad \text{with}\quad
   \widehat{R}(u)=
  \begin{pmatrix}
        1&0&0&0 \\
        0&b(u)&c(u)&0 \\
        0&c(u)&b(u)&0 \\
        0&0&0&1
  \end{pmatrix},
\enq
\noindent and
\beq
  b(u)=\frac{\sinh u}{\sinh (u +\eta )},\quad
  c(u)=\frac{\sinh \eta }{\sinh (u +\eta )}.
\enq
The bulk
monodromy matrix $T(\la)\in\mathrm{End}(V_0\otimes\EuScript{H})$, $V_0\simeq\Cset^2$, is defined as an ordered product of $R$ matrices:
\beq
T_0\pa{\la}=R_{0M}\pa{\la-\xi_M} \dots R_{01}\pa{\la-\xi_1}
=\pa{\ba{cc}
                    A\pa{\la}&B\pa{\la}\\
                    C\pa{\la}&D\pa{\la}
                    \ea}_{[0]} \,\, .
\enq
The subscript $0$ labels here the two-dimensional auxiliary
space $V_0$, whereas subscripts $m$ running from $1$ to $M$ refer to the
quantum spaces $\EuScript{H}_m$ of the chain. Besides, we attach an
inhomogeneity parameter $\xi_m$ to each site $m$ of the chain. We
recall that $T\pa{\la}$ satisfies the Yang-Baxter algebra, on $V_0\otimes V_{0'}\otimes\EuScript{H}$:
\beq
R_{00'}\pa{\la-\mu} T_{0}\pa{\la} T_{0'}\pa{\mu}= T_{0}\pa{\la}
T_{0'}\pa{\mu} R_{00'}\pa{\la-\mu}.
\enq

Let us also introduce the two boundary matrices,
$K_{\pm}\pa{\la}=K\pa{\la\pm\tf{\eta}{2};\xi_{\pm}}$,
where $K\pa{\la;\xi}$ is the $2\times 2$ matrix acting on the auxiliary space:
\beq
    K\pa{\la;\xi}= \pa{\ba{cc}
                                \s{\la+\xi}&0\\
                                0&\s{\xi-\la} \ea }_{\pac{0}} \,\, .
\enq
The boundary monodromy matrix $\mc{U}\pa{\la}$ \cite{Skl88} is built out of a
product of $T\pa{\la}$ and $K_+\pa{\la}$, namely\footnote{Note that it corresponds to the matrix $\mathcal{U}_+$ of our previous article (I). Since we consider only the `+' case in the present article, we do not specify it in the notations.}
\beq
\mc{U}^{\e{t}_0}_0=T_0^{\e{t}_0}\pa{\la}K^{\e{t}_0}_+\pa{\la}\widehat{T}_0^{\e{t}_0}\pa{\la}
                    =\pa{\ba{cc}
                    \mc{A}\pa{\la}&\mc{B}\pa{\la}\\
                    \mc{C}\pa{\la}&\mc{D}\pa{\la}
                    \ea}^{t_0}_{[0]} \, \, ,
\label{Monodromy matrix Boundary}
\enq
where
\begin{align}
\widehat{T}_0\pa{\la}
&= R_{10}\pa{\la+\xi_1-\eta} \dots
R_{M0}\pa{\la+\xi_M-\eta} \nonumber \\
&= \pa{-1}^M \pl_{j=1}^{M} [ \s{\la+\xi_j} \s{\la+\xi_j-    \eta} ]\,\,\,
T^{-1}_{0}\pa{-\la +\eta} \,\, .
\end{align}
This boundary monodromy matrix satisfies
the reflection algebra first introduced in \cite{Che84}:
\begin{multline}
R_{0 0'}\pa{-\la+\mu}\,
 \mc{U}^{\e{t}_0}_0\pa{\la}\, R_{0 0'}\pa{-\la-\mu-\eta}\,
 \mc{U}^{\e{t}_{0'}}_{0'}\pa{\mu}\\
=\mc{U}^{\e{t}_{0'}}_{0'}\pa{\mu}\, R_{0 0'}\pa{-\la-\mu-\eta}\,
\mc{U}^{\e{t}_0}_0\pa{\la}\, R_{0 0'}\pa{-\la+\mu}
 \label{algerbe+}.
\end{multline}

The commuting charges of the  XXZ spin-$\tf{1}{2}$ chain with
diagonal boundary conditions are realized by the one-parameter
family of transfer matrices:
\beq
\mc{T}\pa{\la}=\e{tr}_0\pac{\,\,\mc{U}_{0}\pa{\la}K_{-}\pa{\la}}
\,\, ,
\enq
and the Hamiltonian \eqref{HamXXZ} is obtained in terms of the
derivative $\f{\dd \mc{T}\pa{\la}}{\dd \la}\mid_{\la=\tf{\eta}{2}}$
in the homogeneous case $\xi_i = \tf{\eta}{2}$, $i=1,\ldots,M$.

Common eigenstates of all transfer matrices (and thus of the
Hamiltonian \eqref{HamXXZ} in the homogeneous case) can be
constructed by successive actions of $\mc{B}\pa{\la}$ operators on
the reference state $\ket{0}$ which is the ferromagnetic state with
all the spins up. More precisely, the state \footnote{In order to
lighten the formulae we have slightly changed the notation with respect to the one in Paper I. Namely, the vector $\ket{\paa{\la}_1^n}_{b}$
corresponds to $\ket{\psi_+\pa{\paa{\la}}}$ in
\cite{KitKMNST07}. Such a boundary state should in particular be distinguished from the corresponding bulk state that we merely denote $\ket{\paa{\la}_1^n}$.}
\beq
\ket{\paa{\la}_1^n}_{b} \equiv \mc{B}\pa{\la_1}\dots
\mc{B}\pa{\la_n}\ket{0}
\label{definition vecteur Bethe Bord}
 \enq
is a common eigenstate of the transfer matrices if the set of
spectral parameters $\{\la\}_{1 }^{n}\equiv \{\la_j\}_{1\le j\le n}$ is a solution of the Bethe
equations
\beq
 y_j \pa{\la_j;\paa{\la}_1^n}
=y_j\pa{-\la_j;\paa{\la}_1^n},
\qquad j=1,\ldots,n,
\label{Bethe-eq}
 \enq
where
\begin{align}
 &y_j\pa{x;\paa{\la}_1^n}
  = \f{\hat{y}\pa{x;\paa{\la}_1^n}}{\sd{\la_j,x-\eta}} \; ,
\nonumber \\%
 &\hat{y}\pa{x;\paa{\la}_1^n}=
    - a\pa{x}\,d\pa{-x}
\s{x+\xi_+-\tf{\eta}{2}}\s{x+\xi_- -\tf{\eta}{2}}
\nonumber\\
& \hspace{8cm} \times \pl_{l=1}^{n} \sd{x-\eta,\la_l}
   \,\, \label{Yfunction} \; .
\end{align}
Here and in the following, $\sd{\la,\mu}$ denotes the function
\beq
\sd{\la,\mu}=\s{\la+\mu}\s{\la-\mu} \,\, ,
\enq
and the functions $a\pa{\la}$  and  $d\pa{\la}$ stand
respectively for the eigenvalues of the bulk operators $A(\la)$
and $D(\la)$ on the pseudo-vacuum $\ket{0}$:
\beq
 a\pa{\la}=\pl_{i=1}^{M}\s{\la-\xi_i+\eta}, \qquad d\pa{\la}=\pl_{i=1}^{M} \s{\la-\xi_i}.
\enq
Of course it is also possible to implement the Bethe ansatz starting from the
dual state $\bra{0}$ and acting on it with $\mc{C}\pa{\la}$
operators:
\beq
_b\bra{\paa{\la}_1^n} \equiv \bra{0} \mc{C}\pa{\la_1} \dots
\mc{C}\pa{\la_n}.
\enq

The description of the ground state of $\mc{H}$ in the half-infinite
chain depends on the regime. One should distinguish the two domains
$-1<\Delta\le 1$ (massless regime) and $\Delta>1$ (massive regime):
\begin{alignat*}{4}
  &\alpha_j=\la_j,\quad  & &\zeta=i\eta>0,\quad &
  &\xi_-=-i\tilde\xi_-,\quad \text{with}\
  -\frac{\pi}{2}<\tilde\xi_-\le\f{\pi}{2},\qquad  &
  &\text{for}\ -1<\Delta\le 1,\\
  &\alpha_j=i\la_j,\quad  & &\zeta=-\eta>0,\quad &
  &\xi_-=-\tilde\xi_-+i\delta\f{\pi}{2},\quad\text{with}\
  \tilde\xi_-\in\R, \qquad  &
  &\text{for}\ \Delta>1,
\end{alignat*}
where $\delta=1$ for $| h_-|<\sinh\zeta$ and $\delta=0$ otherwise.
Thus, to a given set of roots $\{\la_j\}$ corresponds a set of
variables $\{\a_j\}$ given by the previous change of variables.
Note that the nature of the ground state rapidities depends on the value of
the boundary field $h_-$.

Indeed, when $\tilde\xi_-<0$ or $\tilde\xi_->\zeta/2$, the ground state of the Hamiltonian
\eqref{HamXXZ} is given in both regimes by the maximum number $N$ of
roots $\la_j$ corresponding to real (positive) $\alpha_j$ such that
$\cos p(\la_j) <\Delta$. In the thermodynamic limit $M\tend\infty$,
these roots $\la_j$ form a dense distribution on an interval
$[0,\Lambda]$ of the real or imaginary axis. Their density
\begin{equation}
   \rho(\la_j)=\lim_{M\tend\infty} [M(\la_{j+1}-\la_j)]^{-1}
\end{equation}
satisfies the integral equation
\begin{equation}\label{Liebbulk}
2\pi \rho\pa{\la}+\Int{-\Lambda}{\Lambda} \,\f{i
\sinh (2\eta)}{\sd{\la-\mu,\eta}}\rho\pa{\la}\, \dd \la
        =\f{2i\sinh\eta}{ \sd{\la,\tf{\eta}{2}}},
\end{equation}
with $\Lambda=+\infty$ in the massless regime, and
$\Lambda=-i\pi/2$ in the massive one. The density can be expressed in terms of usual functions:

\beq
\rho\pa{\la}
  = \left\{ \ba{lc}
            \f{1}{\zeta\cosh\pa{\tf{\pi\la}{\zeta}} }, & -1<\Delta<1 ; \\
            \f{i}{\pi} \pl_{n\geq1}^{} \paf{\sinh n\zeta}{\cosh{n\zeta}}^{2}
            \f{\theta_3\pa{i\la;\e{-\zeta}}}{\theta_4\pa{i\la;\e{-\zeta}}},  & 1<\Delta   .
            \ea \right.
\enq

However, when $0<\tilde\xi_-<{\zeta}/{2}$, the
ground state also admits a root $\check\la$ (corresponding to a
complex $\check\alpha$) which tends to $\eta/2-\xi_-$ with
exponentially small corrections in the large $M$ limit. In that
case, the density of real roots is still given by the solution of \eqref{Liebbulk}.

\section{The ABA approach to correlation functions}\label{sec-method}

  A zero temperature correlation
function is the normalized expectation value, in the ground state of the Hamiltonian \eqref{HamXXZ}, of some \textit{local}\footnote{ \textit{i.e}. acting
non-trivially only in $ \tens_{k=1}^{m}{\EuScript{H}}_k$.} operator $\mc{O}_m$,
\beq
\label{Cor-fun}
\moy{\mc{O}_m}= \f{_b\bra{\paa{\la}_1^N} \mc{O}_m
\ket{\paa{\la}_1^N}_b}{_b\bra{\paa{\la}_1^N}\!\! \ket{\paa{\la}_1^N}_b},
\enq
where the parameters $\la$  are the solutions of the ground state Bethe
equations.

In order to compute such a correlation function, one should first  derive
 the action of the corresponding local operator on the boundary state $\ket{\paa{\la}_1^N}_b$,
and then evaluate the resulting scalar products.
 We have constructed in \cite{KitKMNST07} a method to solve this problem.
This method is based on a revisited version of the quantum inverse problem, first introduced in \cite{KitMT99,MaiT00}
for the XXZ spin chain with periodic boundary conditions. Once a local operator
 is reconstructed in terms of the entries of the bulk monodromy matrix, its action on boundary
states can then be computed thanks to the decomposition of boundary states in terms of bulk states and to the Yang-Baxter commutation relations.

We shall now recall the main points of our method.

\subsection{The bulk inverse problem revisited}

\begin{prop}[Solution of the bulk inverse problem]
\label{Problem inverse bulk}
 \cite{KitMT99,MaiT00} Let $E^{ij}_{m}$ be
an elementary matrix acting non-trivially only on the
$\e{\textit{m}}^{\e{th}}$ site of the chain, then
\beq
E^{ij}_{m}= \pl_{k=1}^{m-1} \pa{A+D}\pa{\xi_k}
\; \mathrm{tr}\pa{T_0\pa{\xi_m} E^{ij}_{0}} \; \pl_{k=1}^{m}
\pa{A+D}^{-1}\pa{\xi_k}.
\label{formule du probleme inverse}\enq
\end{prop}

Note that, thanks to the crossing symmetry of the $R$-matrix, one can
recast the inverse of the bulk transfer matrix at inhomogeneity parameter $\pa{A+D}^{-1}\pa{\xi_k}$ in terms of the transfer matrix at shifted parameter
$\pa{A+D}\pa{\xi_k-\eta}$, namely
\beq
\pa{A+D}^{-1}\pa{\xi_k} = \f{\pa{A+D}\pa{\xi_k-\eta}}{a\pa{\xi_k}
d\pa{\xi_k-\eta}} \,\, .
\enq

It is worth pointing out that the products of elementary matrices on the first $m$ sites of the chain define a basis in the space of local operators $\mc{O}_m$, so that
\eqref{formule du probleme inverse} allows one to define a reconstruction for all such operators. However, this reconstruction is especially convenient when one wants to obtain the action on a bulk Bethe state; indeed, in such a case, the product of bulk transfer matrices merely produces a numerical factor.
This is no longer the case when one acts on a boundary Bethe state. The theorem below allows
one to reconstruct local operators in a way adapted to an action on boundary states.

\begin{theorem}
\label{theoremannulationoperteurs}\cite{KitKMNST07}
For any set of inhomogeneity parameters $\{\xi_{i_1},\ldots,\xi_{i_n}\}$, the product of bulk operators
\beq\label{prod-n}
  T_{\eps_{i_n}\, \eps'_{i_n}}(\xi_{i_n})\dots T_{\eps_{i_1}\, \eps'_{i_1}}(\xi_{i_1})\,
  T_{\bar\eps_{i_1}\, \bar\eps'_{i_1}}(\xi_{i_1}-\eta)\dots
       T_{\bar\eps_{i_n}\,\bar\eps'_{i_n}}(\xi_{i_n}-\eta)
\enq
vanishes if, for some  $k\in\{i_1,\ldots,i_n\}$, $\eps_k=\bar\eps_k$.
\end{theorem}

Thus we have :
\begin{cor}\label{cor-elem}
A product of elementary matrices acting on the first $m$ sites of the chain can be expressed as
a single monomial in the entries of the bulk monodromy matrix:
\begin{multline}
  E_{1}^{\eps_1\,\eps'_1}\dots E_{m}^{\eps_m\,\eps'_m}=
  \pl_{i=1}^{m} \big[a(\xi_i)\, d(\xi_i-\eta)\big]^{-1}
               \\
  \times
  T_{\eps'_1\,\eps_1}(\xi_1)\dots T_{\eps'_m\,\eps_m}(\xi_m)\,
  T_{\bar\eps_m\,\bar\eps_m}(\xi_m-\eta) \dots
  T_{\bar\eps_1\,\bar\eps_1}(\xi_1-\eta)
\label{blockselementaires}
\end{multline}
with $\bar\eps_i=\eps'_i+ 1 \pmod 2$.
\end{cor}

This result represents a strong simplification. Indeed, it means that, over the $2^m$ monomials appearing in the
reconstruction of a local operator \eqref{formule du probleme inverse}, only one is non-zero. We shall now explain how to compute the action of this non-vanishing monomial on an arbitrary (bulk or boundary) state.

\subsection{Action on bulk and boundary states}

Before stating the lemma which explains how to derive the action of the former monomial on a bulk state, we recall that the action of $A(\mu)$ or $D(\mu)$ on a bulk state $\ket{\paa{\la}_1^N}\equiv\prod_{j=1}^N B(\la_j)\ket{0}$ produces two kinds of terms: the {\em direct term}, where all rapidities remain
unchanged, and {\em indirect terms } where one $\la_j$ is replaced by $\mu$.

\begin{lemme}[Action on a bulk state]\label{lem-action}\cite{KitKMNST07}
The action  on a bulk state $\ket{\paa{\la}_1^N}$ of a string of operators
\beq
  \mc{O}_{\eps_{i_1},\ldots,\eps_{i_n}}^{\eps'_{i_1},\ldots,\eps'_{i_n}}=
  \underbrace{T_{\eps'_{i_n}\,\eps_{i_n}}(\xi_{i_n})\dots
              T_{\eps'_{i_1}\,\eps_{i_1}}(\xi_{i_1})}_{\pa{1}}
  \underbrace{T_{\bar{\eps}_{i_1}\,\bar{\eps}_{i_1}}(\xi_{i_1}-\eta) \dots
              T_{\bar{\eps}_{i_n}\,\bar{\eps}_{i_n}}(\xi_{i_n}-\eta)}_{\pa{2}}
\enq
with $\bar\eps_l=\eps'_l+1\pmod 2$, satisfies the restrictions:
\begin{itemize}
  \item The only non-zero contributions of the tail operators $\pa{2}$ come from
\begin{itemize}
 \item[(i)]  the indirect action  of all $A(\xi_l-\eta)$ operators;

 \item[(ii)]  the direct action of all $D(\xi_l-\eta)$ operators.
\end{itemize}
\item In what concerns the head operators $\pa{1}$,
\begin{itemize}
  \item[(iii)]  if $\eps'_l=1$, the action of the operator
$T_{\eps'_l\,\eps_l}(\xi_l)$ (i.e. $A(\xi_l)$ or $B(\xi_l)$) does not result in any substitution of a parameter $\xi_i-\eta$;
 \item[(iv)]  if $\eps'_l=2$,  the action of the operator
$T_{\eps'_l\,\eps_l}(\xi_l)$ (i.e. $D(\xi_l)$ or $C(\xi_l)$) substitutes $\xi_l-\eta$ with $\xi_l$; moreover,
if there were others parameters $\xi_j-\eta$, $j\ne l$, in the initial state, they are still present in the resulting state.
\end{itemize}
\end{itemize}
\end{lemme}

This lemma enables us to compute the action of local operators on any (arbitrary) bulk state. In order to compute the action on a boundary state, we use the fact that the latter can be decomposed in terms of bulk states:

\begin{prop}[Boundary-bulk decomposition]\label{prop-bound-bulk-decomp}\cite{KitKMNST07},
\cite{Wan02} Let  $\ket{\paa{\la}_1^n}_b$ be an arbitrary boundary state, then it can be expressed in terms of bulk states as
\beq
 \ket{\paa{\la}_1^n}_b
    =\sum_{\substack{\sg_{i}=\pm \\ i=1,\ldots,n }}
     H_{\paa{\sg_i}}^{\mc{B}}\pa{\paa{\la}_1^n}
       \ket{\paa{\la^{\sg}}_1^n} ,
   \label{boundary bulk decomposition}
\enq
with
\begin{equation}\label{boundary-bulk_coeff}
 H_{\paa{\sg_i}}^{\mc{B}}\pa{\paa{\la}_1^n}
    =\pl_{j=1}^{n} H_{\sg_j}^{\mc{B}}(\la_j)\ \cdot \!\!
     \pl_{1\leq r<s\leq n}\!\!\f{\sinh(\ov{\la}^{\sg}_{rs}-\eta)}{\sinh (\ov{\la}^{\sg}_{rs})}.
\end{equation}
In this expression, $H_{\sg}^{\mc{B}}(\la)$ denotes the ``one-particle'' boundary-bulk coefficient, which can be written as
\begin{equation}\label{boundary-bulk_1p}
 H_{\sg}^{\mc{B}}(\la)=\sigma \, (-1)^M d(-\la^{\sg})\, \f{\s{2\la+\eta}}{\sinh 2\la}\, \s{\la^\sg+\xi_{+}-\tf{\eta}{2}}.
\end{equation}
%
%
Here we have used the notations:
\beqa
&&\la_{rs} = \la_r - \la_s, \qquad \ov{\la}_{rs}=\la_r+\la_s, \\
&&\la_j^{\sg_j}= \sg_j \la_j,  \quad \e{and} \,\, \e{more} \,\, \e{generally} \quad
\paa{\la_{}^{\sg_{}}}_1^{n}= \paa{\la_j^{\sg_j}}_1^{n} \,\, .
\eeqa
\end{prop}

It is remarkable that, by using this decomposition and the previous lemma, we are able to express the action of a local operator $\mc{O}_m$ on an arbitrary boundary state as a linear
combination of such boundary states:
\beq
\mc{O}_m \ket{\paa{\la}_1^N}_b=\sul_{\alpha_m} C_{\alpha_m}(\paa{\la};\paa{\xi}) \,
\ket{\paa{\mu_i}_{i\in \alpha_m}}_b,
\label{action generique sur etat bord d'un operateur local}
 \enq
where the summation is taken over certain subsets $\paa{\mu_i}_{i\in \alpha_m}$ of $\paa{\la}_1^N\cup\paa{\xi}_1^m$, and where $C_{\alpha_m}$ are coefficients which can be computed generically\footnote{ See section 5.3 of \cite{KitKMNST07} for the explicit expression in the case of a product of elementary matrices.}.

\subsection{From scalar products to correlation functions}

It now remains, in order to obtain the correlation function \eqref{Cor-fun}, to take the scalar product of this resulting combination of states with the ground state ${}_b\bra{\paa{\la}_1^N}$. This can be done by using the trigonometric generalization \cite{KitKMNST07} of the rational \cite{Wan02} formula for the scalar product between a boundary Bethe state and an arbitrary boundary state. In particular, we have to evaluate the following type of renormalized scalar product:
\beq
\mathbb{S}\pa{\paa{\la},\paa{\mu}}=
\f{_b\bra{\paa{\la}}\!\!\ket{\paa{\mu}}_b}
{_b\bra{\paa{\la}}\!\!\ket{\paa{\la}}_b},
\enq
where the sets $\paa{\la}$ and $\paa{\mu}$ are partitioned
according to:
\beq
 \paa{\la}=\paa{\la_a}_{a\in\alpha_-}\!\cup\paa{\la_b}_{b\in\alpha_+}\,\, ,\qquad
 \paa{\mu}=\paa{\la_a}_{a\in\alpha_-}\!\cup\paa{\xi_{b}}_{b\in\gamma_+} \,\, ,
\enq
with $|\alpha_+|=|\gamma_+|$. Here the parameters $\la$ are the solutions of the ground state boundary
Bethe equations, $\paa{\xi_{b}}_{b\in \gamma_+}$ are arbitrary
inhomogeneities and $\alpha_+\cup\alpha_-$ is a partition of $\{1,\ldots,N\}$.

Since we are especially interested in the thermodynamic limit $M \tend +\infty$ of the
correlation function \eqref{Cor-fun}, we only recall the explicit formula for the leading
asymptotic contribution of the renormalized scalar product \cite{KitKMNST07}:
\begin{equation}
 \mathbb{S}\pa{\paa{\la},\paa{\mu}}=\paf{1}{M}^{\abs{\a_+}}
  \EuScript{S} (\{\la\}_{\alpha_+},\{\xi\}_{\gamma_+};\{\la\}_{\alpha_-}) \
   \underset{\substack{a\in \a_+\\b\in\gamma_+}}{\text{det}}\big[\widetilde{\mc{S}}_{ab}\big].
\label{Produit scalaire reduit}
\end{equation}

The coefficient $\EuScript{S} (\{\la\}_{\alpha_+},\{\xi\}_{\gamma_+};\{\la\}_{\alpha_-})$ has been computed in \cite{KitKMNST07}. Note that, since $\{\la\}$ is a solution of the boundary Bethe equations, there is some sign arbitrariness in the expression of this coefficient: indeed, it is in fact equal to
\begin{multline}\label{coeff-prod-sc}
  \EuScript{S}_{\sg} (\{\la\}_{\alpha_+},\{\xi\}_{\gamma_+};\{\la\}_{\alpha_-})
  =  \f{\pl_{\substack{a,b\in\alpha_+ \\ a>b}} \sd{\la_b,\la_a}}
       {\pl_{\substack{a,b\in\gamma_+ \\ a>b}}\sd{\xi_{b},\xi_{a}}}\,
 \pl_{a\in\alpha_-}
  \f{\pl_{b\in\alpha_+}\sd{\la_{b},\la_a}}
    {\pl_{b\in\gamma_+}\sd{\xi_{b},\la_a}}\\
 \times\prod_{b\in\gamma_+}\frac{\hat{y}(\xi_{b};\{\la\}_{\alpha_+\cup\alpha_-})\,
                              \sinh (2\xi_{b}+\eta) }{\sinh(2\xi_b)}
    \prod_{a\in\alpha_+}\frac{\sinh(2\la_a^\sg-\eta)\,\sinh(2\la_a)}
       {\hat{y}(\la_{a}^\sg;\{\la\}_{\alpha_+\cup\alpha_-})\, \sinh (2\la_{a}+\eta) },
\end{multline}
for any value of $\sg_a \in\{+,-\}$, $a\in\alpha_+$,
where $\hat{y}$ is the function defined in
\eqref{Yfunction}.
%
%

When $M$ is large, the matrix elements of $\widetilde{\mathcal{S}}$ reduce to
\beq
  \widetilde{ \mc{S}}_{ab}
   \underset{M\tend\infty}{\sim}
       \left\{\ba{c c}
          2i\pi M \s{\la_a-\xi_-+\eta/2} \Psi\pa{\la_a,\xi_{b}}
            &\e{ if}\ \la_a=\check \la \; ,\smallskip \\
           \rho^{-1}\pa{\la_a} \, \Psi\pa{\la_a,\xi_{b}}            &\e{ if}\ \la_a \ne \check \la\; ,
       \ea \right.
\enq
the corrections being of order $\e{O}\pa{\tf{1}{M}}$, and
\beq
\Psi\pa{\la,\xi}= \f{\rho\pa{\la-\xi}-\rho\pa{\la-\eta+\xi}}{2 \s{2\xi-\eta}} \; .
\enq

The determinant structure of the scalar product as well as
peculiarities of the coefficients $C_{\alpha_m}$ enable us to write:
\beq
\moy{\mc{O}_m} = \f{1}{M^m} \sul_{\substack{\paa{\nu}_{I}\subset
\paa{\la}_1^{N}\cup\paa{\xi}_1^m \\ \abs{I}=m}}
H_m\pa{\paa{\nu}_{I},\paa{\xi}_1^m} \pa{1+\e{O}\pa{\tf{1}{M}}},
\enq
in which the coefficient $H_m\pa{\paa{\nu}_{I},\paa{\xi}_1^m}$ can be computed generically.
Taking the thermodynamic limit $M\tend+\infty$ we recast the sums over replaced rapidities $\la$
into integrals:
\beq
\f{1}{M}\sul_{i=1}^{N} \sum_{\sg_i=\pm} \sg_i\, f\pa{\la_i^\sg} \underset{M \tend
+\infty}{\longrightarrow} \Int{\mathscr{C}}{} \dd \la \,\, \rho\pa{\la}
f\pa{\la}, \hspace{1cm} \forall f\in \mc{C}^{0}\pa{\mathscr{C}}.
\enq
In the boundary model the contour of integration $\mathscr{C}$ depends on the anisotropy
parameter $\Delta$ and on the boundary field $h_-$.

\section{Action of local operators on boundary states}
\label{Action of local operators on boundary states}

Using Corollary~\ref{cor-elem}, Lemma~\ref{lem-action} and the boundary-bulk decomposition of Proposition~\ref{prop-bound-bulk-decomp}, it is easy to compute the action of a product of elementary matrices of the form~\eqref{blockselementaires} on an arbitrary boundary state. This computation was explicitely performed in \cite{KitKMNST07}, and enabled us there to obtain some expressions for the elementary building blocks of correlation functions.

The aim of the present article is to obtain such expressions for physical correlation functions, and in particular for one-point functions. Therefore, if we want to use the method recalled in Section~\ref{sec-method}, the main problem is to obtain some resummed formulas directly for the action of the local spin operators we consider. This is the purpose of the present section.

In the first part of this section, we derive the action of the operator
\beq
\mc{Q}_m\pa{\kappa}\equiv \pl_{i=1}^{m}
\pa{E_{i}^{11}+\kappa E_{i}^{22}}=
\pl_{i=1}^{m}\pa{A+\kappa
D}\pa{\xi_i}\pl_{i=1}^{m}\pa{A+D}^{-1}\pa{\xi_i}
\label{Q kappa definition}
 \enq
on arbitrary boundary states. $\moy{\mc{Q}_m\pa{\kappa}}$ can be interpreted
in the boundary model as the generating function of the magnetization
at a distance $m$ from the boundary:
\begin{equation}
\moy{\frac{1-\sg_m^z}{2}}=D_m \partial_{\kappa} \moy{\mc{Q}_m\pa{\kappa}}|_{\kappa=1},
\end{equation}
where $D_m$  is the lattice derivative : $D_m \, u_m \equiv u_{m+1}-u_m$.

Then, in the second part of this section, we give the formulas for the action of the local spin operators $E_m^{22}=\frac{1-\sg_m^z}{2}$, $E_m^{12}=\sg_m^+$
and $E_m^{21}=\sg_m^-$ on arbitrary boundary states. Note that the action
of $E_m^{11}$ follows from the fact that $E_m^{11}=1-E^{22}_m$.

\subsection{Action of $\mc{Q}_m\pa{\kappa}$}

We start by computing the action of $\mc{Q}_m\pa{\kappa}$ on an arbitrary
bulk state, and then infer from this formula its action on arbitrary boundary states.

\begin{prop}
\label{Theorem Qkappa}
 The action of
$\mc{Q}_m\pa{\kappa}$ on an arbitrary bulk state
$\ket{\paa{\la}_1^N}$ can be expressed as
\beq
\mc{Q}_m\pa{\kappa}\ket{\paa{\la}_1^N}=\sul_{n=0}^{m}
\sul_{\mc{P}_{\la} \,;\,  \mc{P}_{\xi} }
 R^{\kappa}_n\pa{\mc{P}_{\la} \, , \,  \mc{P}_{\xi}} \,\,
\ket{\paa{\xi}_{\gamma_+} \!\! \cup \paa{\la}_{\a_-}}\; .
\label{action Q(kappa) sur bulk}
\enq
In the above formula, we sum over all possible partitions $\mc{P}_{\la}$ and $\mc{P}_{\xi}$ of the sets
$\paa{\la}_1^N \!\!\!$ and $\paa{\xi}_1^m \!\!$ into subsets $\paa{\la}_{\a_+}
\!\!\cup \paa{\la}_{\a_-}$ and $\paa{\xi}_{\gamma_+}\!\!\cup
\paa{\xi}_{\gamma_-}$ respectively, satisfying the constraint on the cardinality
$\abs{\a_+}=\abs{\gamma_+}=n$:
\begin{alignat}{3}
&\mc{P}_{\la} & &:\; \paa{\la}_1^N=\paa{\la}_{\a_+}\cup \paa{\la}_{\a_-} ,
          & \quad & \abs{\a_+}=n \, , \\
&\mc{P}_{\xi} & &:\; \paa{\xi}_1^m=\paa{\xi}_{\gamma_+}\cup \paa{\xi}_{\gamma_-},
          & \quad  & \abs{\gamma_+}=n \,
\label{partition des xi} \, .
\end{alignat}
The coefficient $R^{\kappa}_n\pa{\mc{P}_{\la} \, ,\,
\mc{P}_{\xi}}$ splits into two parts,
\beqa\label{def-Rn^kappa}
R^{\kappa}_n\pa{\mc{P}_{\la} \, ,\,  \mc{P}_{\xi}}=
R\pa{\mc{P}_{\la} \, ,\, \mc{P}_{\xi}}\;
S^{\kappa}_n\big(\paa{\xi}_{\gamma_+},\paa{\la}_{\a_+}\big) \, ,
\label{definition de R_n^Kappa}
\eeqa
the first one having a product
structure,
\beq
R\pa{\mc{P}_{\la} \, , \, \mc{P}_{\xi}}
= \f{\pl_{a\in\a_+} \bigg\{ a\pa{\la_a}\pl_{b\in \a_-}f\pa{\la_b,\la_a}\bigg\}}
{\pl_{a\in\gamma_+} \bigg\{a\pa{\xi_a} \pl_{b\in\a_-\cup\a_+} f\pa{\la_b,\xi_a}\bigg\}}
\,\, \pl_{a\in \gamma_-} \f{\pl_{b\in \gamma_+} f\pa{\xi_b,\xi_a}}
{\pl_{b\in \a_+} f\pa{\la_b,\xi_a}},
\label{definition du facteur R_n}
\enq
and the second one, which depends here only on the subsets $\paa{\la}_{\a_+}$ and $\paa{\xi}_{\gamma_+}$, being given as a ratio of two determinants,
\beq
S^{\kappa}_n\pa{\paa{\nu}_1^{n},\paa{\mu}_1^n}=
\det{n}{M_\kappa\big(\paa{\mu}_1^n,\paa{\nu}_1^{n}\big)}\
\mathrm{det}_{n}^{-1}\pa{\f{1}{\s{\nu_k-\mu_j+\eta}}}
\,\, . \label{definition de S_n}
\enq
The entries of the matrix $M_\kappa$ read
\beq
 \big[M_\kappa\big(\paa{\mu}_1^n,\paa{\nu}_1^{n}\big)\big]_{jk}=
t\pa{\nu_k,\mu_j}-\kappa \, t\pa{\mu_j,\nu_k} \pl_{\substack{a=1 \\
a\not= j}}^{n} \f{f\pa{\mu_a,\mu_j}}{f\pa{\mu_j,\mu_a}}
\pl_{a=1}^{n} \f{f\pa{\mu_j,\nu_a}}{f\pa{\nu_a,\mu_j}} \;\; ,
\label{definition de Mjk(kappa)}
\enq
and the functions $f$ and $t$ stand for
\beq
t\pa{\la,\mu} =  \f{\sinh\eta}{\s{\la-\mu}\s{\la-\mu+\eta}}  , \qquad
f\pa{\la,\mu} = \f{\s{\la-\mu + \eta}}{\s{\la-\mu}}.
\enq
\end{prop}


The above theorem appears as a non-trivial generalization of
the action of $\mc{Q}_m\pa{\kappa}$ on bulk Bethe eigenvectors \cite{KitMST02a}.
Indeed, when $\ket{\paa{\la}}$ is not an eigenstate of the bulk
transfer matrix, then $\prod_{i=1}^{m}\pa{A+D}\pa{\xi_i}$
does not act by multiplication any more. Of course our result reproduces the previous
 case when we send the parameters $\la$ to a solution of
the bulk Bethe equations.

\Proof
The proof goes by induction on $m$.

Property \eqref{action Q(kappa) sur bulk} is obvious for $m=1$.
Assume that it holds for some $m$. To prove its validity for $m+1$ we have to compute
\beq
\mc{Q}_{m+1}\pa{\kappa}\ket{\paa{\la}_1^N}=\f{\pa{A+\kappa D}\pa{\xi_{m+1}}\mc{Q}_m\pa{\kappa}
\pa{A+D}\pa{\xi_{m+1}-\eta}}{a(\xi_{m+1})d(\xi_{m+1}-\eta)} \ket{\paa{\la}_1^N}.
\enq

Let us first reproduce the coefficient $R^{\kappa}_n\pa{\mc{P}_{\la} \, , \, \mc{P}_{\xi}}$ in the case when the partition $\mc{P}_{\xi}$ is such that $\xi_{m+1} \not\in \paa{\xi}_{\gamma_+}$. The corresponding state
$\ket{\paa{\la}_{\a_-}\cup\paa{\xi}_{\gamma_+}}$ can only be obtained by the direct action
of $\pa{A+\kappa D}\pa{\xi_{m+1}}$. In order to reproduce the claimed form of the coefficient $R_{n}^{\kappa}$
it is enough to prove that  $\pa{A+\kappa D}\pa{\xi_{m+1}-\eta}$ acts directly. Suppose that this is not the case.
Then $\mc{Q}_m\pa{\kappa}$ acts on a state containing $\xi_{m+1}-\eta$. In virtue of Lemma~\ref{lem-action}, the action of $\mc{Q}_m\pa{\kappa}$ on these states cannot replace $\xi_{m+1}-\eta$. Thus $\pa{A+D}\pa{\xi_{m+1}}$ exchanges $\xi_{m+1}-\eta$ with $\xi_{m+1}$, which leads to a contradiction.

We still have to reproduce the coefficient $R^{\kappa}_n\pa{\mc{P}_{\la} \, , \, \mc{P}_{\xi}}$ corresponding to states $\ket{\paa{\la}_{\a_-}\cup\paa{\xi}_{\gamma_+}}$ such that $\xi_{m+1}\in\paa{\xi}_{\gamma_+}$. Theorem \ref{theoremannulationoperteurs} yields the decomposition:
\begin{equation*}
\mc{Q}_{m+1}\pa{\kappa}=\f{A\pa{\xi_{m+1}}\mc{Q}_m\pa{\kappa}}{a(\xi_{m+1})d(\xi_{m+1}-\eta)}
\underbrace{D\pa{\xi_{m+1}-\eta}}_{\pa{1}}
+ \f{\kappa D\pa{\xi_{m+1}}\mc{Q}_m\pa{\kappa}}{a(\xi_{m+1})d(\xi_{m+1}-\eta)}
\underbrace{A\pa{\xi_{m+1}-\eta}}_{\pa{2}} \; ,
\end{equation*}
whereas Lemma \ref{lem-action} ensures that
\begin{itemize}
\item $\pa{1}$ only acts directly; indeed $A\pa{\xi_{m+1}}$ cannot replace
$\xi_{m+1}-\eta$ by $\xi_{m+1}$;
\item $\pa{2}$ acts indirectly and thus $D\pa{\xi_{m+1}}$ only acts by
substitution.
\end{itemize}
The  formula for $R_n^{\kappa}$ \eqref{definition de R_n^Kappa} follows after computing the resulting actions and rearranging the sums thanks to the re-summation formula
provided by the contour integral:
\beq
0=\oint\limits_{\mathbb{R}\cup\mathbb{R}+i\pi}
 \f{\dd z}{\s{z-\xi_{n+1}}}\pl_{a=1}^{r}\f{f\pa{z,x_a}}{f\pa{\xi_{n+1},x_a}}\
 S^{\kappa}_{n+1}\big(\paa{\xi}_1^{n}\cup\paa{z},\paa{\la}_1^{n+1}\big) \;\; .
\label{identite de resommation}
\enq
Note that the parameters $x_a$ appearing in
the contour integral \eqref{identite de resommation} are generic.
\qed

Using the boundary-bulk decomposition of Proposition~\ref{prop-bound-bulk-decomp}, one can now deduce from Proposition~\ref{Theorem Qkappa} the action of
$\mc{Q}_m\pa{\kappa}$ on arbitrary boundary states.

\begin{cor}\label{Cor Qkappa}
The action of $\mc{Q}_m\pa{\kappa}$ on an arbitrary boundary state
$\ket{\paa{\la}_1^N}_b$ reads:
\beq
\mc{Q}_m\pa{\kappa}\ket{\paa{\la}_1^N}_b=\sul_{n=0}^{m}
\sul_{\substack{\mc{P}_{\la} \,;\,  \mc{P}_{\xi} }}
 \mc{R}^{\kappa}_{n}\pa{\mc{P}_{\la},\mc{P}_{\xi}} \,\,
\ket{\paa{\xi}_{\gamma_+} \!\! \cup \paa{\la}_{\a_-}}_b.
\label{action Q(kappa) sur bord}
 \enq
The sum over partitions is defined as in Theorem
\ref{Theorem Qkappa}, and the coefficient
$\mc{R}^{\kappa}_n$ can be expressed as
\beq\label{def-coeff-bord}
\mc{R}^{\kappa}_{n}\pa{\mc{P}_{\la},\mc{P}_{\xi}}=
\sul_{ \substack{\sg_i=\pm \\ i  \in\a_+}}
\mc{R}_{\sigma}\pa{\mc{P}_{\la},\mc{P}_{\xi}}\;
S^{\kappa}_n (\paa{\xi}_{\gamma_+}\!\! ,\paa{\la^{\sg}}_{\a_+} ) \, ,
\enq
where  $S^{\kappa}_n\pa{\paa{\nu}_1^n,\paa{\mu}_1^n}$ is the bulk function
defined in \eqref{definition de S_n}, while $\mc{R}_{\sigma}\pa{\mc{P}_{\la},\mc{P}_{\xi}}$ is the boundary dressing of \eqref{definition du facteur R_n}:
\begin{multline}
\mathcal{R}_{\sigma }(\mathcal{P}_{\lambda },\mathcal{P}_{\xi })
 =\frac{\prod\limits_{a\in\alpha _{+}}\bigg\{
        a(\lambda _{a}^{\sigma })\prod\limits_{b\in \alpha_{-}}
        \big[ f(\lambda _{b},\lambda _{a}^{\sigma })\,
              f(-\lambda _{b},\lambda_{a}^{\sigma })\big]\bigg\} }
       {\prod\limits_{a\in \gamma _{+}}\bigg\{
        a(\xi_{a})\prod\limits_{b\in \alpha _{+}}f(\lambda _{b}^{\sigma },\xi_{a})
        \prod\limits_{b\in \alpha _{-}}
        \big[f(\lambda _{b},\xi _{a})f(-\lambda_{b},\xi _{a})\big]\bigg\} } \\
  \times
  \prod\limits_{a\in \gamma _{-}}
  \frac{\prod\limits_{b\in \gamma_{+}} f(\xi _{b},\xi _{a})}
       {\prod\limits_{b\in \alpha _{+}}f(\lambda_{b}^{\sigma },\xi _{a})}\
  \frac{H_{\{\sigma \}_{\alpha _{+}}}^{\mathcal{B}}(\{\lambda \}_{\alpha_{+}})}
       {H^{\mathcal{B}}(\{\xi \}_{\gamma _{+}})}.\label{def-Rn-bord}
\end{multline}
Here $H_{\{\sigma \}_{\alpha _{+}}}^{\mathcal{B}}(\{\lambda \}_{\alpha_{+}})$ and $H^{\mathcal{B}}(\{\xi \}_{\gamma _{+}})$ stand for the boundary-bulk coefficients~\eqref{boundary-bulk_coeff} associated respectively to $\{\la\}_{\alpha_+}$, $\{\sigma\}_{\alpha_+}$, and to  $\{\xi\}_{\gamma_+}$, $\{\sigma\}_{\gamma_+}=\{1,\ldots,1\}$.
\end{cor}

\Proof The proof is a straightforward consequence of the boundary-bulk decomposition
\eqref{boundary bulk decomposition} applied to Proposition~\ref{Theorem Qkappa}.
More precisely, expressing the boundary state $\ket{\{\la\}_1^N}_b$ in terms of the bulk states $\ket{\{\la^\sg\}_1^N}$, and using~\eqref{action Q(kappa) sur bulk}, we get
\begin{equation*}
 \mc{Q}_m\pa{\kappa}\ket{\paa{\la}_1^N}_b=\sul_{n=0}^{m}
\sul_{\substack{\mc{P}_{\la} \,;\,  \mc{P}_{\xi} }}
\sul_{ \substack{\sg_i=\pm \\ 1\le i  \le N}}
 H_{\{\sigma \}}^{\mathcal{B}}(\{\lambda \}_1^N)\;
 R^{\kappa}_n\pa{\mc{P}_{\la^\sg} \, , \,  \mc{P}_{\xi}} \,\,
\ket{\paa{\xi}_{\gamma_+} \!\! \cup \paa{\la^\sg}_{\a_-}}.
\end{equation*}
We now use the fact that
\begin{equation*}
  H_{\{\sigma \}}^{\mathcal{B}}(\{\lambda \}_1^N)
  =\pl_{b\in\alpha_-}\frac{\pl_{a\in\alpha_+}f(-\la_b^\sg,\la_a^\sg)}
                          {\pl_{a\in\gamma_+}f(-\la_b^\sg,\xi_a)}\
   \frac{H_{\{\sigma \}_{\alpha _{+}}}^{\mathcal{B}}(\{\lambda \}_{\alpha_{+}})}
        {H^{\mathcal{B}}(\{\xi \}_{\gamma _{+}})}\
   H_{1,\{\sigma \}_{\alpha _{-}}}^{\mathcal{B}}(\{\xi\}_{\gamma_+}\cup\{\lambda \}_{\alpha_{-}}),
\end{equation*}
where $H_{1,\{\sigma \}_{\alpha _{-}}}^{\mathcal{B}}(\{\xi\}_{\gamma_+}\cup\{\lambda \}_{\alpha_{-}})$ is the boundary-bulk coefficient of $\ket{\{\xi\}_{\gamma_+}\cup\{\lambda \}_{\alpha_{-}}}_b$ in terms of $\ket{\{\xi\}_{\gamma_+}\cup\{\lambda^\sg \}_{\alpha_{-}}}$.
Note that the first factor of this product combines with the products over $b\in\alpha_-$ in the expression \eqref{definition du facteur R_n} of $R(\mc{P}_{\la^\sg},\mc{P}_\xi)$, and that the resulting factor,
\begin{equation*}
 \pl_{b\in\alpha_-}
 \frac{\pl_{a\in\alpha_+}\big[f(\la_b^\sg,\la_a^\sg)\,f(-\la_b^\sg,\la_a^\sg)\big]}
      {\pl_{a\in\gamma_+}\big[f(\la_b^\sg,\xi_a)\,f(-\la_b^\sg,\xi_a)\big]},
\end{equation*}
is actually independant of the value of $\sg_i$ for $i\in\alpha_-$. It enables us to reconstruct the boundary state $\ket{\{\xi\}_{\gamma_+}\cup\{\lambda \}_{\alpha_{-}}}_b$, with a coefficient which reduces to~\eqref{def-coeff-bord}.
\qed

\subsection{Action of local spin operators}
\label{action operateurs locaux sur Bord}

We list here the action of the local spin operators $\sg_m^-$, $\sg_m^+$ and $E_m^{22}$ on bulk and boundary states. We omit the proofs since, although a
little more technical, they parallel the one concerning the action
of $\mc{Q}_m\pa{\kappa}$.

\begin{prop}\label{prop-act-spin}
 The action of $\sg_m^-$, $E_m^{22}$ and $\sg_m^+$ on an arbitrary bulk state $\ket{\{\la\}_1^N}$ can be expressed as
\begin{align*}
 &\sg_m^-\; \ket{\{\la\}_1^N}= \sum_{n=0}^{m-1} \sum_{\mathcal{P}_\la^-,\mathcal{P}_\xi}
   R_n^-(\mathcal{P}_\la^-,\mathcal{P}_\xi)\
    \ket{\{\xi\}_{\gamma_+}\cup\{\la\}_{\alpha_-}},
            \displaybreak[0]\\
 &E_m^{22}\; \ket{\{\la\}_1^N}= \sum_{n=0}^{m-1} \sum_{c_1=1}^N
    \sum_{\mathcal{P}_\la^{22},\mathcal{P}_\xi}
   R_n^{22}(\mathcal{P}_\la^{22},\mathcal{P}_\xi)\
    \ket{\{\xi\}_{\gamma_+}\cup\{\la\}_{\alpha_-}},
            \displaybreak[0]\\
 &\sg_m^+\; \ket{\{\la\}_1^N}= \lim_{\la_{N+1}\tend\xi_m} \sum_{n=0}^{m-1}
   \sum_{c_1=1}^N \sum_{\substack{c_2=1\\ c_2\ne c_1}}^{N+1}
   \sum_{\mathcal{P}_\la^+,\mathcal{P}_\xi}
   R_n^+(\mathcal{P}_\la^+,\mathcal{P}_\xi)\
  \ket{\paa{\xi}_{\gamma_+} \cup \paa{\la}_1^N \setminus \paa{\la}_{\widetilde{\a}_+}},
\end{align*}
in which the sums run over the following partitions
\begin{alignat}{3}
&\mc{P}_{\xi} & &:\; \paa{\xi}_1^{m}=\paa{\xi}_{\gamma_+}\cup \paa{\xi}_{\gamma_-},
          & \quad  \text{with}\quad & \abs{\gamma_+}=n+1 ,\displaybreak[0]\\
&\mc{P}^{-}_{\la} & &:\; \paa{\la}_1^{N}=\paa{\la}_{\a_+}\cup \paa{\la}_{\a_-} ,
          & \quad \text{with}\quad & \abs{\a_+}=n , \\
&\mc{P}^{22}_{\la} & &:\;
      \paa{\la_k}_{\substack{1\le k\le N\\ k\ne c_1}}
  =\paa{\la}_{\a_+}\cup \paa{\la}_{\a_-} ,
          & \quad \text{with}\quad & \abs{\a_+}=n , \label{partition_lambda22}\\
&\mc{P}^+_{\la} & &:\;
  \paa{\la_k}_{\substack{1\le k\le N\\ k\ne c_1,c_2}}
  =\paa{\la}_{\a_+}\cup \paa{\la}_{\a_-} ,
          & \quad \text{with}\quad & \abs{\a_+}=n . \label{partition_lambda+}\displaybreak[0]\\
\intertext{We also define the following partitions, associated respectively to \eqref{partition_lambda22} and to \eqref{partition_lambda+},}
&\widetilde{\mc{P}}^{22}_{\la} & &:\;
  \paa{\la}_{1}^{N}
  =\paa{\la}_{\widetilde{\a}_+}\cup \paa{\la}_{{\a}_-} ,
          & \quad \text{with}\quad & \widetilde{\a}_+=\alpha_+\cup\{c_1\},
                 \label{partition_lambda22tilde}\\
&\widetilde{\mc{P}}^+_{\la} & &:\;
  \paa{\la}_{1}^{N+1}
  =\paa{\la}_{\widetilde{\a}_+}\cup \paa{\la}_{\widetilde{\a}_-} ,
          & \quad \text{with}\quad & \widetilde{\a}_+=\alpha_+\cup\{c_1,c_2\} .
                 \label{partition_lambda+tilde}
\end{alignat}

The coefficients $R_n^-(\mathcal{P}^-_\la,\mathcal{P}_\xi)$,   $R_n^{22}(\mathcal{P}^{22}_\la,\mathcal{P}_\xi)$ and $R_n^+(\mathcal{P}^+_\la,\mathcal{P}_\xi)$ are given as
\begin{align}
 &R_n^-(\mathcal{P}_\la^-,\mathcal{P}_\xi)= R(\mathcal{P}_\la^-,\mathcal{P}_\xi)\
  \lim_{\xi\tend\xi_m}
  \frac{\pl_{a\in\gamma_+}\sinh(\xi_a-\xi)}{\pl_{a\in\alpha_+}\sinh(\la_a-\xi)}\
 \widehat{S}_n(\{\la\}_{\alpha_+},\{\xi\}_{\gamma_+};\xi,\emptyset),
          \displaybreak[0]\\
 &R_n^{22}(\mathcal{P}_\la^{22},\mathcal{P}_\xi)
      = R(\widetilde{\mc{P}}_\la^{22},\mathcal{P}_\xi)\
  \sinh \eta  \pl_{a\in\alpha_+} f(\la_a,\la_{c_1})\nonumber\\
 &\hspace{1.5cm}\times
  \lim_{\xi\tend\xi_m}
  \frac{\pl_{a\in\gamma_+}\sinh(\xi_a-\xi)}{\pl_{a\in\widetilde{\alpha}_+}\sinh(\la_a-\xi)}\
 \widehat{S}_n(\{\la\}_{\alpha_+},\{\xi\}_{\gamma_+};\xi,\{\la_{c_1}\}),
          \displaybreak[0]\\
 &R_n^{+}\big(\mc{P}^+_{\la} \, ,\, \mc{P}_{\xi}\big) =
R \big( \widetilde{\mc{P}}^+_{\la} \, , \, \mc{P}_{\xi} \big)\
 f(\la_{c_2},\la_{c_1})\;
 \pl_{i=1}^{2}  \bigg\{\sinh \eta  \pl_{a\in\alpha_+} f(\la_a,\la_{c_i})\, \bigg\}
 \nonumber\\
 &\hspace{1.5cm}\times
 \frac{\pl_{a\in\gamma_+}\sinh (\la_{N+1}-\xi_a+\eta)}
      {\pl_{a\in\widetilde{\alpha}_+} \sinh (\la_{N+1}-\la_a+\eta)}\
\widehat{S}_n\big(\paa{\xi}_{\gamma_+},\paa{\la}_{\a_+}; \la_{N+1},\{\la_{c_1},\la_{c_2}\}\big).
\end{align}
Here $R(\mathcal{P}_\la,\mathcal{P}_\xi)$ is given by \eqref{definition du facteur R_n}, and the structure of the factor $\widehat{S}_n(\{\xi\}_1^{n+1},\{\la\}_1^n;\xi,\{\mu\}_1^p)$ is similar to \eqref{definition de S_n}:
\begin{multline}\label{def-Sn^theta}
 \widehat{S}_n(\{\xi\}_1^{n+1},\{\la\}_1^n;\xi,\{\mu\}_1^p)=
 \frac{\pl_{a=1}^n\pl_{b=1}^{n+1}\sinh(\xi_b-\la_a+\eta)}
      {\pl_{a>b}\sinh(\xi_a-\xi_b)\pl_{a>b}\sinh(\la_b-\la_a)}\\
\times
 \det{n+1}{\widehat{M}(\{\la\}_1^n,\{\xi\}_1^{n+1};\xi,\{\mu\}_1^p)},
\end{multline}
where the matrix elements of $\widehat{M}$ are obtained as
\begin{equation}\label{Mhat}
 \big[\widehat{M}\big(\{\la\}_1^n,\{\xi\}_1^{n+1};\la_{n+1},\{\mu\}_1^p\big)\big]_{jk}=
 \big[M_{\kappa(\la_j,\{\mu\})}\big(\paa{\la}_1^{n+1},\paa{\xi}_1^{n+1}\big) \big]_{jk} ,
\end{equation}
with
\beq
 \kappa(\la_j,\{\mu\})=\pa{1-\delta_{j,n+1}}
        \pl_{i=1}^{n'}\f{f\pa{\mu_i,\la_j}}{f\pa{\la_j,\mu_i}} \nonumber  ,
\enq
in which $\delta_{ij}$ denotes the Kronecker symbol and $M_\kappa$ is defined as in \eqref{definition de Mjk(kappa)}.
\end{prop}

Using again the boundary-bulk decomposition, we are now in position to list the action of local spin operators on boundary states.

\begin{cor}
With the same notations as in Proposition~\ref{prop-act-spin}, the action of $\sg_m^-$, $E_m^{22}$ and $\sg_m^+$ on an arbitrary boundary state  $\ket{\{\la\}_1^N}_b$ takes
the form
\begin{align*}
 &\sg_m^-\; \ket{\{\la\}_1^N}_b= \sum_{n=0}^{m-1} \sum_{\mathcal{P}_\la^-,\mathcal{P}_\xi}
   \mc{R}_n^-(\mathcal{P}_\la^-,\mathcal{P}_\xi)\
    \ket{\{\xi\}_{\gamma_+}\cup\{\la\}_{\alpha_-}}_b,
           \displaybreak[0]\\
 &E_m^{22}\; \ket{\{\la\}_1^N}_b= \sum_{n=0}^{m-1} \sum_{c_1=1}^N
    \sum_{\mathcal{P}_\la^{22},\mathcal{P}_\xi}
   \mc{R}_n^{22}(\mathcal{P}_\la^{22},\mathcal{P}_\xi)\
    \ket{\{\xi\}_{\gamma_+}\cup\{\la\}_{\alpha_-}}_b,
           \displaybreak[0]\\
 &\sg_m^+\; \ket{\{\la\}_1^N}_b= \lim_{\la_{N+1}\tend\xi_m} \sum_{n=0}^{m-1}
   \sum_{c_1=1}^N \sum_{\substack{c_2=1\\ c_2\ne c_1}}^{N+1}
   \sum_{\mathcal{P}_\la^+,\mathcal{P}_\xi}
   \mc{R}_n^+(\mathcal{P}_\la^+,\mathcal{P}_\xi)\
  \ket{\paa{\xi}_{\gamma_+} \cup \paa{\la}_1^N \setminus \paa{\la}_{\widetilde{\a}_+}}_b.
\end{align*}
The boundary coefficients $\mc{R}^-$,  $\mc{R}^{22}$ and  $\mc{R}^+$ have a structure similar to their
corresponding bulk counterparts:
\begin{align*}
 &\mc{R}_n^-(\mathcal{P}_\la^-,\mathcal{P}_\xi)
  = \sul_{ \substack{\sg_i=\pm \\ i \in \a_+ }}
    \mc{R}_{\sg}(\mathcal{P}_\la^-,\mathcal{P}_\xi)\
  \lim_{\xi\tend\xi_m}
  \frac{\pl_{a\in\gamma_+}\sinh(\xi_a-\xi)}{\pl_{a\in\alpha_+}\sinh(\la_a-\xi)}\
 \widehat{S}_n(\{\la\}_{\alpha_+},\{\xi\}_{\gamma_+};\xi,\emptyset),
           \displaybreak[0]\\
 &\mc{R}_n^{22}(\mathcal{P}_\la^{22},\mathcal{P}_\xi)
      = \sul_{ \substack{\sg_i=\pm \\ i \in \widetilde{\a}_+ }}
  \mc{R}_\sg (\widetilde{\mc{P}}_\la^{22},\mathcal{P}_\xi)\
  \sinh \eta  \pl_{a\in\alpha_+} f(\la_a,\la_{c_1})\nonumber\\
 &\hspace{3cm}\times
  \lim_{\xi\tend\xi_m}
  \frac{\pl_{a\in\gamma_+}\sinh(\xi_a-\xi)}{\pl_{a\in\widetilde{\alpha}_+}\sinh(\la_a-\xi)}\
 \widehat{S}_n(\{\la\}_{\alpha_+},\{\xi\}_{\gamma_+};\xi,\{\la_{c_1}\}),
        \displaybreak[0]\\
 &\mc{R}^{+}_{n}\big(\mc{P}^+_{\la}, \mc{P}_{\xi}\big) =
   \sul_{ \substack{\sg_i=\pm \\ i \in \widetilde{\a}_+ }}
   \mc{R}_{\sg}\big(\widetilde{\mc{P}}^+_{\la}, \mc{P}_{\xi}\big)\
 f\big(\la_{c_2}^\sg,\la_{c_1}^\sg\big)\;
 \pl_{i=1}^{2}  \bigg\{\sinh \eta
                      \pl_{a\in\alpha_+} f\big(\la_a^\sg,\la_{c_i}^\sg\big)\, \bigg\}
 \nonumber\\
 &\times
 \frac{\pl_{a\in\gamma_+} \big[ f(-\la_{N+1},\xi_a)\,\sinh(\la_{N+1}-\xi_a+\eta)\big]}
      {\pl_{a\in\widetilde{\alpha}_+}
         \big[ f(-\la_{N+1},\la_a^\sg)\,\sinh(\la_{N+1}-\la_a^\sg+\eta) \big]}\
\widehat{S}_n\big(\paa{\xi}_{\gamma_+},\paa{\la^\sg}_{\a_+}; \la_{N+1},\{\la_{c_i}^\sg\}\big),
\end{align*}
where $\mc{R}_{\sg}$ is defined as in \eqref{def-Rn-bord} and $\widehat{S}_n$ is the bulk quantity~\eqref{def-Sn^theta}.
\end{cor}

\section{Correlation functions in the half-infinite chain}
\label{Q kappa a limite M infini}

We apply the results of the previous section to derive the expectation values
of the generating function $\moy{\mc{Q}_m\pa{\kappa}}$ of $\moy{\sg_m^z}$, and of  $\moy{\sg_1^+ \sg_{m+1}^-}$ in the ground state of the half-infinite chain. These are
the boundary analogues of the results published in
\cite{KitMST02a}.

\subsection{The generating function $\moy{\mc{Q}_m\pa{\kappa}}$}

\begin{prop}
The generating function  $\moy{\mc{Q}_m\pa{\kappa}}$ is obtained, in the thermodynamic limit $M\tend+\infty$, as the homogeneous limit of the quantity
\begin{align}
\moy{\mc{Q}_m\pa{\kappa}}
 =& \sul_{n=0}^{m} \f{1}{\pa{n!}^2}
    \oint\limits_{\Gamma_+\pa{\paa{\xi}_1^m}} \!\! \f{\dd^n z}{\pa{2i\pi}^n}
    \Int{\mc{C}_{D}}{} \dd^n\la \
    \pl_{a=1}^m \pl_{b=1}^{n}\f{f\pa{z_b,\xi_a}}{f\pa{\la_b,\xi_a}}\
    \mathcal{W}_-(\paa{\la}_1^n,\paa{z}_1^n)
         \nonumber \\
&\hspace{3cm}\times
  \det{n}{{M}_\kappa\pa{\paa{\la},\paa{z}}}\ \det{n}{\Psi\pa{\la_j,z_k}},
\label{Q kappa original series}
\end{align}
in which  ${M}_\kappa$ is given by \eqref{definition de Mjk(kappa)}, and $\mathcal{W}_-$ is the boundary dressing,
\begin{multline}\label{W-bord}
 \mathcal{W}_-\big(\{\la\}_1^{n_1},\{ z\}_1^{n_2}\big)
 =\f{\pl_{j=1}^{n_2}\s{z_j+\xi_- - \tf{\eta}{2}}}
    {\pl_{j=1}^{n_1}\s{\la_j+\xi_- - \tf{\eta}{2}}}\\
  \times
  \frac{\pl_{a=1}^{n_1}\pl_{b=1}^{n_2}\s{z_b+\la_a-\eta}}
       {\pl_{a<b}^{n_2}\s{\ov{z}_{ab}-\eta}\pl_{a<b}^{n_1}\s{\ov{\la}_{ab}-\eta}}\
  W\big(\paa{\la}_1^{n_1},\paa{z}_1^{n_2}\big),
\end{multline}
of the bulk quantity
\begin{equation}\label{W-bulk}
 W\big(\paa{\la}_1^{n_1},\paa{z}_1^{n_2}\big)
 =\f{\pl_{a=1}^{n_1}\pl_{b=1}^{n_2}\big[ \s{z_b-\la_a-\eta}\,\s{z_b-\la_a+\eta}\big]}
    {\pl_{a,b=1}^{n_1}\s{\la_{ab}-\eta}\pl_{a,b=1}^{n_2}\s{z_{ab}+\eta}}.
\end{equation}
The contour of integration $\mc{C}_D$ depends on the boundary magnetic field $h_-$:
\beq\label{contour_D}
\mc{C}_D=\left\{  \ba{ll}
         \intoo{-\Lambda}{\Lambda} \cup\Gamma_+\pa{\check{\la}}\quad
                                        & \text{if}\quad 0<\tilde{\xi}_-<\tf{\zeta}{2}, \\
         \intoo{-\Lambda}{\Lambda} & \text{otherwise},
         \ea\right.
\enq
where $\Gamma_{\pm}\pa{z}$ stands for a small loop of index $\pm 1$ with respect to $z$. We recall that
$\Lambda =+\infty$ for $-1<\Delta<1$ and $\Lambda=-i\tf{\pi}{2}$ for $\Delta>1$.
\end{prop}

\Proof
Corollary \ref{Cor Qkappa} yields the action of
$\mc{Q}_m\pa{\kappa}$ on a boundary state. It is convenient to note that the coefficient
$\mc{R}_{\sg}\big(\mc{P}_{\la},\mc{P}_{\xi}\big)$ \eqref{def-Rn-bord} can be rewritten as
\begin{align}
 \mc{R}_{\sg}\big(\mc{P}_{\la},\mc{P}_{\xi}\big)
 &=\bigg(\pl_{a\in\alpha_+}\sg_a\bigg)\ (\sinh \eta)^{|\gamma_+|}\!\! \pl_{b\in\gamma_+\cup\gamma_-}
  \frac{\pl_{\substack{a\in\gamma_+\\ a\ne b}} f(\xi_a,\xi_b)}
       {\pl_{a\in\alpha_+} f(\la_a^\sg,\xi_b)}\quad
  \mathcal{W}_-(\{\la^\sg\}_{\alpha_+},\{\xi\}_{\gamma_+})
   \nonumber\\
  &\hspace{0cm}\times
  \frac{\pl_{a>b}\sinh(\xi_a-\xi_b)\pl_{a>b}\sinh(\la_b-\la_a)}
       {\pl_{a\in\alpha_+}\pl_{b\in\gamma_+}\sinh(\xi_b-\la_a+\eta)}\
  {\EuScript{S}}_{\sg}(\{\la\}_{\alpha_+},\{\xi\}_{\gamma_+};\{\la\}_{\alpha_-})^{-1},
\end{align}
in which $\EuScript{S}_{\sg}(\{\la\}_{\alpha_+},\{\xi\}_{\gamma_+};\{\la\}_{\alpha_-}) $ is the function defined in \eqref{coeff-prod-sc}.
Then, using the reduced scalar
product formula \eqref{Produit scalaire reduit} and absorbing the sums
over partitions $\mc{P}_{\xi}$ into auxiliary $z$ integrals\footnote{We refer the
reader to \cite{KitMST02a} for technical details.}, we obtain
the former representation.

 Note that the contour contains $\Gamma_+\pa{\check\la}$ for large positive
 boundary field since we have to absorb the contribution coming
 from the replacement of the complex root $\check\la$ as explained in
 \cite{KitKMNST07}.
\qed

\subsection{The ground state expectation value $\moy{\sg_1^+\sg_{m+1}^-}$}

Using the same method as for the generating function $\moy{\mc{Q}_m(\kappa)}$, we can also compute the ground state expectation value $\moy{\sg_1^+\sg_{m+1}^-}$. It gives
\begin{align}
 \moy{\sg_1^+\sg_{m+1}^-}
 &= \sum_{n=0}^{m-1}\f{\sinh (\xi_1+\xi_--\eta/2)}{n!(n+1)!} \!\!
    \oint\limits_{\Gamma_+ (\{\xi\}_1^{m+1})}\!\! \pl_{k=1}^{n+1} \frac{\dd z_k}{2i\pi}\
    \Int{\mc{C}_D}{}\pl_{k=1}^{n+1} \dd\la_k\ \Int{\mc{C}_A}{} \dd\la_{n+2}
                   \nonumber\\
 &\times
    \pl_{a=2}^{m+1}\f{\pl_{b=1}^{n+1}f\pa{z_b,\xi_a}}{\pl_{b=1}^{n}f\pa{\la_b,\xi_a}}\
    \pl_{b=1}^{n}\f{\sinh(\la_b-\xi_1)}{\sinh(\la_b-\xi_{m+1})}
    \pl_{b=1}^{n+1}\f{\sinh(z_b-\xi_{m+1})}{\sinh(z_b-\xi_{1})}
          \nonumber\\
 &\times
    \f{\pl_{b=1}^{n+1}\sinh(\la_b-\la_{n+1}+\eta)\pl_{b=1}^{n+2}\sinh(\la_{n+2}-\la_b+\eta)}
      {\pl_{b=1}^{n+1}\big[\sinh(z_b-\la_{n+1}+\eta)\,\sinh(\la_{n+2}-z_b+\eta)\big]}\
    \mathcal{W}_-\big(\paa{\la}_1^{n+2},\paa{z}_1^{n+1}\big)
         \nonumber\\
 &\times
    \f{\pl_{b=1}^{n+2}\sinh(\xi_1+\la_b-\eta)}{\pl_{b=1}^{n+1}\sinh(\xi_1+z_b-\eta)}\
  \det{n+1}{\widehat{M}\big(\paa{\la}_1^{n},\paa{z}_1^{n+1};\xi_{m+1}\big)}
         \nonumber\\
 &\times
  \det{n+2}{\Psi(\la_{j},\xi_1),\Psi(\la_{j},z_{1}),\dots ,\Psi (\lambda_{j},z_{n+1})} .
\end{align}
In this expression, $\mc{W}_-$ denotes the boundary quantity \eqref{W-bord}, $\widehat{M}\big(\paa{\la}_1^{n},\paa{z}_1^{n+1};\xi_{m+1}\big)$ is a simplified notation for the matrix $\widehat{M}\big(\paa{\la}_1^{n},\paa{z}_1^{n+1};\xi_{m+1},\emptyset\big)$ defined in \eqref{Mhat}, $\mc{C}_D$ is the contour \eqref{contour_D}, and $\mc{C}_A$ denotes the following contour ($A$-type contour):
\begin{equation}\label{contour-A}
{\mc{C}}_A=\left\{  \ba{ll}
            \intoo{-\Lambda+\eta}{\Lambda+\eta} \cup\Gamma_-\pa{\check{\la}},
              &\text{if}\quad -\tf{\zeta}{2}<\tilde{\xi}_-<0, \\
            \intoo{-\Lambda+\eta}{\Lambda+\eta}, & \text{otherwise.}
            \ea\right.
\end{equation}
In the homogeneous limit, this results simplifies into
\begin{multline}\label{sp-sm-hom}
 \moy{\sg_1^+\sg_{m+1}^-}
 = \sum_{n=0}^{m-1}\f{\sinh \xi_-}{n!(n+1)!} \!\!
    \oint\limits_{\Gamma_+ (\eta/2)}\!\! \pl_{k=1}^{n+1} \frac{\dd z_k}{2i\pi}\
    \Int{\mc{C}_D}{}\pl_{k=1}^{n+1} \dd\la_k\ \Int{\mc{C}_A}{} \dd\la_{n+2}
                   \nonumber\\
 \quad\times
   \pl_{a=1}^{n+1} \bigg[ \f{\sinh (z_{a}+\tf{\eta }{2})}
                            {\sinh (z_{a}-\tf{\eta }{2})}  \bigg]^m
    \pl_{a=1}^{n} \bigg[ \f{\sinh(\la_{a}-\tf{\eta }{2})}
                           {\sinh(\la_{a}+\tf{\eta }{2})} \bigg]^m \
    \f{\pl_{b=1}^{n+2}\sinh(\la_b-\eta/2)}{\pl_{b=1}^{n+1}\sinh(z_b-\eta/2)}
                    \nonumber\\
 \quad\times
    \f{\pl_{b=1}^{n+1}\sinh(\la_b-\la_{n+1}+\eta)\pl_{b=1}^{n+2}\sinh(\la_{n+2}-\la_b+\eta)}
      {\pl_{b=1}^{n+1}\big[\sinh(z_b-\la_{n+1}+\eta)\,\sinh(\la_{n+2}-z_b+\eta)\big]}\
    \mathcal{W}_-\big(\paa{\la}_1^{n+2},\paa{z}_1^{n+1}\big)
         \nonumber\\
 \quad\times
  \det{n+1}{\widehat{M}(\paa{\la}_1^{n},\paa{z}_1^{n+1};\eta/2)}
  \det{n+2}{\Psi(\la_{j},\eta/2),\Psi(\la_{j},z_{1}),\dots ,\Psi (\lambda_{j},z_{n+1})} ,
\end{multline}
%
%
%
%


\section{An alternative resummation}
\label{Resommation alternative}

\subsection{Bulk type resumations}

We have obtained in the previous sections a series representation
for the generating function. It happens, just as in the bulk case
\cite{KitMST04c}, that it is also possible to derive
a totally different representation for $\moy{ \mc{Q}_m\pa{\kappa}}$. The latter is based
on a re-summation of its expansion
with respect to elementary blocks :
\beq
\moy{ \mc{Q}_m\pa{\kappa}}=\moy{\pl_{i=1}^{m}
\pa{E_{i}^{11}+\kappa E_{i}^{22}}} = \sul_{s=0}^m \kappa^s F_s
\,\, ,
\enq
where
\beq
F_s=\f{1}{s!\,\pa{m-s}!}\sul_{\pi\in\Sigma_m}\langle
E^{\eps_{\pi\pa{1}}\,\eps_{\pi\pa{1}}}_{1}\dots
E^{\eps_{\pi\pa{m}}\,\eps_{\pi\pa{m}}}_{m}\rangle,
 \quad
 \eps_i= \left\{\ba{c l } 2,& i=1...s,\\
              1, & i=s+1...m,
                 \ea\right.
\label{sommesurpermuattion Fs}
\enq
and $\Sigma_m$ is the group of permutations of $m$ elements. These elementary blocks were computed in \cite{KitKMNST07}. They can be written as multiple
 integrals in the half-infinite size limit:
\beqa
&&\moy{ E_{1}^{\eps_1\,\eps'_1} \dots E_{m}^{\eps_m\,\eps'_m}}
 =
\pa{-1}^{m-s} \Int{\mc{C}_{D}}{} \pl_{i=1}^{s}\dd \la_i
 \Int{\mc{C}_{A}}{}  \pl_{i=s+1}^{m} \!\! \dd \la_{i}
\ \f{\text{det}_m\pac{\Psi\pa{\la_i,\xi_j}}
 } {\pl_{i<j}^m\big[\sinh \xi_{ij}\,\s{\overline{\xi}_{ij}-\eta}\big]}
 \nonumber \\
&& \hspace{-2mm}\times\f{\pl_{i,j}^m \s{\la_i+\xi_j-\eta}}{\pl_{i>j}
\s{\la_{ij}-\eta} \s{\ov{\la}_{ij}-\eta}}
\; \pl_{p=1}^s\pac{\pl_{j=1}^{i_p-1}\s{\xi_j-\la_{p}}\pl_{j=i_p+1}^{m}\s{\xi_j-\la_p-\eta}}
\nonumber\\
&& \hspace{-2mm}
\times \pl_{i=1}^m\f{\s{\xi_i+\xi_-
-\tf{\eta}{2}}}{\s{\la_{i}+\xi_- -\tf{\eta}{2}}}
\pl_{p=s+1}^{m}\pac{\pl_{j=1}^{i_p-1}\s{\xi_j-\la_p}\pl_{j=i_p+1}^{m}\s{\xi_j-\la_{p}+\eta}} \hspace{-1mm}.
\eeqa
\noindent The indices $i_p$ are defined by
\beq
\left\{\ba{c c c}
        \paa{i\, :\, 1\le i \le m,\, \eps'_i=2}&=&\paa{i_1<\dots<i_s}, \\
        \paa{i\, :\, 1\le i \le m,\, \eps_i=1}&=&\paa{i_{s+1}>\dots>i_m}.
\ea\right.
\enq

For simplicity, we consider from now on the massless regime (although all what follows can be performed in the massive regime as well). In that case, $\eta=-i\zeta$ and the contours of integration $\mc{C}_D$ and $\mc{C}_A$ depend on the boundary
magnetic field $h_-$ as follows:

\vspace{-0.4 cm}

\beq
        \ba{|c |c |c |}
\hline
\e{range}\,\e{of} \,\, \xi_- & {D}-\e{contour} & {A}-\e{contour}  \\
\hline   \tf{\zeta}{2}< \abs{\tilde{\xi}_-}<\tf{\pi}{2} &
\mc{C}_D=\R &
  \mc{C}_A=\R-i\zeta \\
\hline  \tf{\zeta}{2}>\tilde{\xi}_->0 &
\mc{C}^{}_D=\R\bigcup \Gamma_+\pa{\check\la}
 & \mc{C}_A=\R-i\zeta \\
\hline -\tf{\zeta}{2}<\tilde{\xi}_-<0&
\mc{C}^{}_D=\R &
 \mc{C}_A=\paa{\R-i\zeta} \bigcup
\Gamma_-\pa{\check\la} \\
\hline  \ea
\label{contours d'integration}
 \enq
We recall that $\check\la=\tf{\eta}{2}-\xi_-$,   and that
$\Gamma_{\pm}(z) $ is a small loop around $z$ of index $\pm 1$.

We now perform a change of variables in the $A$-type contours:
  $\la'_A = \la_A-i\zeta$. Moreover, we shift the inhomogeneities around  zero
$\delta_i=\xi_i+i\tf{\zeta}{2}$, and define

\beq
a_i=\tf{3}{2}-\eps_i=\left\{ \ba{c c}
                                \tf{1}{2} &  (\eps_i=1) \quad \text{for $A$-type}, \\
                                -\tf{1}{2} & (\eps_i=2) \quad \text{for $D$-type}.
\ea \right.
\enq
This gives
\begin{align}
&\moy{E^{\eps_{\pi\pa{1}}\,\eps_{\pi\pa{1}}}_{1}\dots
E^{\eps_{\pi\pa{m}}\,\eps_{\pi\pa{m}}}_{m}} =
\pa{-1}^{\pac{\pi}} \Int{\mc{C}_D}{} \dd^s \la
\Int{\widetilde{\mc{C}}_A}{} \dd^{m-s} \la\
\f{\det{m}{\widetilde{\Psi}\pa{\la_i,\delta_j}}}{\pl_{i<j}\sd{\delta_i,\delta_j}}
 \nonumber \\
& \hspace{2cm} \times \pl_{j>k}\f{
\s{\delta_k-\la_{\pi\pa{j}}+i a_{\pi\pa{j}}\zeta}
\s{\delta_j-\la_{\pi\pa{k}}-ia_{\pi\pa{k}}\zeta}}
{\s{\la_{\pi\pa{j}\pi\pa{k}}-i\ov{a}_{\pi\pa{j}\pi\pa{k}}\zeta}
\s{\ov{\la}_{\pi\pa{j}\pi\pa{k}}-i\ov{a}_{\pi\pa{j}\pi\pa{k}}\zeta}}
\nonumber \\
& \hspace{2cm} \times \pl_{j,k=1}^{m}\s{\la_j+\delta_k-i a_j\zeta}
\pl_{j=1}^{m}\f{\s{\xi_-+\delta_j}}{\s{\la_j+\xi_--i a_j\zeta}} \;\; .
\label{blockelementaireavecpermutation}
\end{align}
Here
\beq
\widetilde{\Psi}\pa{\la,\delta}=\Psi\pa{\la,\delta-i\tf{\zeta}{2}}=
\f{\rho\pa{\la-\delta}-\rho\pa{\la+\delta}}{2\sinh {2\delta}},
\enq
\begin{align}
&\widetilde{\mc{C}}_A
   =\left\{  \ba{ll}
     \intoo{-\Lambda}{\Lambda} \cup\Gamma_-\pa{\check{\la}-\eta}
                  &\text{if}\ -\tf{\zeta}{2}< \tilde{\xi}_-<0, \\
     \intoo{-\Lambda}{\Lambda} & \text{otherwise.}
    \ea\right.
\end{align}
\noindent and $\pa{-1}^{\pac{\pi}}$ is the signature of the permutation.

One can compute the sum over permutations \eqref{sommesurpermuattion Fs} just as in the bulk case \cite{KitMST04c}.
It leads to the following integral representation for $F_s$:

\begin{prop}[Bulk-type resummation]\label{prop-bulk-type-resum}
The generating function of the spin correlation function $\moy{ \mc{Q}_m\pa{\kappa}}$ can be expressed as
\begin{equation}
 \moy{ \mc{Q}_m\pa{\kappa}}= \sul_{s=0}^m \kappa^s F_s
\end{equation}
with
\begin{multline}
F_s=\f{1}{s!\pa{m-s}!}\Int{\mc{C}_D}{} \dd^s \la
\Int{\widetilde{\mc{C}}_A}{} \dd^{m-s} \la\
\f{\det{m}{\widetilde{\Psi}\pa{\la_i,\delta_j}}}{\pl_{i<j}\sd{\delta_i,\delta_j}}
\,\,
\pl_{j=1}^{m}\f{\s{\xi_-+\delta_j}}{\s{\la_j+\xi_- -ia_j\zeta}}
  \\
  \times  \theta_s\pa{\paa{\la}} \,\,  Z_m\pa{\paa{\la},\paa{\delta}}
\pl_{j,k=1}^{m} \s{\la_j+\delta_k-i\zeta a_j}.
\label{Fs type Bulk}
 \end{multline}
Here $\text{Z}_m\pa{\paa{\la},\paa{\xi}}$ stands for the partition function
of the six-vertex model with domain wall boundary conditions:
\beq
\text{Z}_m\pa{\paa{\la},\paa{\delta}}=
\f{\pl_{j,k}
\sd{\la_j-\delta_k,i\tf{\zeta}{2}}}{\pl_{j<k}\sinh \la_{jk} \sinh \delta_{kj}}\
\det{m}{\f{1}{\sd{\la_j-\delta_k,i\tf{\zeta}{2}}}} \,\,\, ,
\enq
\noindent while
\beq
\theta_s\pa{\paa{\la}}=\pl_{j>k} \f{\sinh \la_{jk}}
{\sd{\la_{jk},i\ov{a}_{jk} \zeta}\s{\ov{\la}_{jk}-i\ov{a}_{jk} \zeta} } \;\; .
\enq
\end{prop}

Similar representations can be obtained for other correlation functions. Here we give only two important examples: the local density of energy and the  $\moy{\sg_{m+1}^+\sg_{1}^-}$ two-point function.

The local density of energy
\beq
E_m=\moy{\sigma^x_{m}\sigma^x_{m+1}+\sigma^y_{m}\sigma^y_{m+1} +\Delta (\sigma^z_{m}\sigma^z_{m+1}-1) }\; ,
\enq
can be written as a sum of $m$ terms
\begin{equation}
 E_m= \sul_{s=0}^{m-1}  \tilde{E}_s,
\end{equation}
each of them containing $m+1$ integrals
\begin{align}
 \tilde{E}_s=&\f{1}{s!\pa{m-1-s}!}\Int{\mc{C}_D}{}\! \dd^{s} \la\!
\Int{\widetilde{\mc{C}}_A}{} \!\dd^{m-s-1} \la\!
\Int{\mc{C}_D}{} \dd \!\la_m\!
\Int{\widetilde{\mc{C}}_A}{}\! \dd \la_{m+1}
\f{\det{m+1}{\widetilde{\Psi}\pa{\la_i,\delta_j}}}{\pl_{i<j}\sd{\delta_i,\delta_j}}
 \nonumber\\ \times&
%
 %
%
   \theta_s\pa{\paa{\la_1,\dots,\la_{m-1}}} \,  Z_m\pa{\paa{\la_1,\dots,\la_{m-1}},\paa{\delta_1,\dots,\delta_{m-1}}}\nonumber\\ \times &
   \pl_{j=1}^{m+1}\f{\s{\xi_-+\delta_j}}{\s{\la_j+\xi_- -ia_j\zeta}}
\pl_{j=m}^{m+1}\pl_{k=1}^{m-1} \f{\sinh(\la_j-\delta_k-ia_j\zeta )\sinh(\la_k-\delta_j+ia_k\zeta )}
{\sinh(\la_{jk}+i\ov{a}_{jk} \zeta)\sinh(\ov{\la}_{jk}-i\ov{a}_{jk} \zeta) }\nonumber\\ \times &
\pl_{j,k=1}^{m+1} \s{\la_j+\delta_k-i a_j\zeta}
\f{\varphi(\la_m,\la_{m+1},\delta_m,\delta_{m+1})}{\sinh(\la_{m+1}-\la_m)\sinh(\la_m+\la_{m+1})}
,
\label{Es type Bulk}
 \end{align}
where
\begin{align*}
\varphi(\la_m,\la_{m+1},\delta_m,\delta_{m+1})=&\sinh(\la_m-\delta_{m+1}+i\frac \zeta 2)\sinh(\la_{m+1}-\delta_{m+1}-i\frac \zeta 2)\\ +&
\sinh(\la_m-\delta_{m}-i\frac \zeta 2)\sinh(\la_{m+1} -\delta_{m} +i\frac \zeta 2)
\\-&\cos\zeta
\sinh(\la_m-\delta_{m+1}-i\frac \zeta 2)\sinh(\la_{m+1}-\delta_{m}-i\frac \zeta 2)
\\-&\cos\zeta
\sinh(\la_m-\delta_{m}+i\frac \zeta 2)\sinh(\la_{m+1}-\delta_{m+1}+i\frac \zeta 2).
\end{align*}

A similar representation can be obtained for the two-point function $\moy{\sg_{m+1}^+\sg_{1}^-}$, namely:
\begin{equation}
 \moy{\sg_{m+1}^+\sg_{1}^-}= \sul_{s=0}^{m-1}  G_s,
\end{equation}
and every term contains $m+1$ integrals
\begin{align}
 G_s=&\f{(-1)^{m-1}}{s!\pa{m-1-s}!}\Int{\mc{C}_D}{} \dd^{s+1} \la
\Int{\widetilde{\mc{C}}_A}{} \dd^{m-s} \la
%
\f{\det{m+1}{\widetilde{\Psi}\pa{\la_i,\delta_j}}}{\pl_{i<j}\sd{\delta_i,\delta_j}}
\nonumber \\ \times&
%
 %
%
   \theta_s\pa{\paa{\la_2,\dots,\la_{m}}} \,  Z_m\pa{\paa{\la_2,\dots,\la_{m}},\paa{\delta_2,\dots,\delta_{m}}}\nonumber\\ \times &
   \pl_{j=1}^{m+1}\f{\s{\xi_-+\delta_j}\sinh(\la_j-\delta_1-ia_j\zeta )\sinh(\la_j-\delta_{m+1}+ia_j\zeta )}{\s{\la_j+\xi_- -ia_j\zeta}}\nonumber\\ \times &\pl_{j,k=1}^{m+1} \s{\la_j+\delta_k-i a_j\zeta}
\pl_{j=1,m+1}\pl_{k=2}^{m} \f{\sinh(\la_j-\delta_k-ia_j\zeta )}
{\sinh(\la_{jk}+i\ov{a}_{jk} \zeta)\sinh(\ov{\la}_{jk}-i\ov{a}_{jk} \zeta) }\nonumber\\ \times&
\f{\sinh(\la_1-\delta_{1}-i\frac \zeta 2)\sinh(\la_{m+1}-\delta_{1}+i\frac \zeta 2)}{\sinh(\la_{m+1}-\la_1)\sinh(\la_1+\la_{m+1})}
,
\label{Gs type Bulk}
 \end{align}

\subsection{Boundary type resumations}

It is important to note that the function $Z_m$ appearing in the representations
\eqref{Fs type Bulk}, \eqref{Es type Bulk}  and \eqref{Gs type Bulk}  is the bulk partition function represented in terms of the Izergin determinant \cite{Ize87}.
In the boundary case one can symmetrize the integrand even further
by writing it in a form invariant under the reversal of the parameters $\la$ and finally rewrite the result in terms of the boundary partition function and the Tsuchiya determinant.
The integration contours in \eqref{Fs type Bulk} are not invariant under the transformation
$\la \tend -\la$. We thus deform the contours until we obtain a
reversal invariant contour. As we do not cross
any pole of the integrand, the result remains unchanged. Actually
we can even pick the contours so as to integrate only over one
contour $\mc{C}$, although this is not necessary. This
contour $\mc{C}$ is defined as follows according to the value of the boundary field $h_-$:
\beq
        \ba{|c|c|c|}
\hline \tf{\zeta}{2}< \abs{\tilde{\xi}_-}<\tf{\pi}{2} &\mc{C}=\R \\
\hline  \tf{\zeta}{2}>\tilde{\xi}_->0 & \mc{C}=\R \,\,\bigcup\,\, \Gamma_{+}(\check\la)
\bigcup\,\, \Gamma_{-}(- \check\la)\\
\hline   -\tf{\zeta}{2}<\tilde{\xi}_-<0  & \mc{C}= \R\,\,
\,\,\bigcup\,\, \Gamma_{+}(i\zeta+\check\la)
\bigcup\,\, \Gamma_{-}(-i\zeta -\check\la) \\
\hline
        \ea
\enq
We extract the totally even part of the integrand appearing in
\eqref{blockelementaireavecpermutation} according to
\beq
\Int{\mc{C}}{} \dd x f\pa{x}=\f{1}{2} \sul_{\sg=\pm}\Int{\mc{C}}{} \dd x
f\pa{x^\sg}\,\, , \qquad x^\sg=\sg \,x \,\, .
\enq

 \noindent We get
\begin{multline}
F_s=\f{1}{s! (m-s)! 2^m}\Int{\mc{C}}{} \dd^m \la\
\f{\det{m}{\widetilde{\Psi}\pa{\la_i,\delta_j}}}{\pl_{i<j}\sd{\delta_i,\delta_j}}
\,\,
\pl_{j=1}^{m}\f{\s{\xi_-+\delta_j}}{\s{\la_j,\xi_- -ia_j\zeta}}
  \\
 \times
 \Theta_s\pa{\paa{\la}} \,\,  H_s\pa{\paa{\la},\paa{\delta}} \;\; .
\label{definition Fs}
\end{multline}
\noindent Here
\beq
\Theta_s\pa{\paa{\la}}=
\pl_{j>k}\f{\sd{\la_j,\la_k}}
{\sd{\la_{jk},i\ov{a}_{jk}\zeta}\sd{\ov{\la}_{jk},i\ov{a}_{jk}\zeta}} \;\; ,
\enq
and the sums over negations have been absorbed into $H_s\pa{\paa{\la},\paa{\delta}}$:
\begin{multline}
H_s\pa{\paa{\la},\paa{\delta}}=\sul_{\sg_i=\pm}
\pl_{j=1}^{m}\big[ \sg_j\,\s{\la^{\sg}_j-\xi_- - i a_j\zeta}\big]
\pl_{j>k}\f{\s{\ov{\la}^{\sg}_{jk}+i\ov{a}_{jk}\zeta}}
{\s{\ov{\la}^{\sg}_{jk}}} \hspace{1cm}  \\
 \times \pl_{j,k}^{m} \s{\la_j^{\sg}+\delta_k-ia_j\zeta}\
\text{Z}_m\pa{\paa{\la^{\sg}},\paa{\delta}}.
\label{Hspresquetsuchida}
\end{multline}

\noindent Equation \eqref{Hspresquetsuchida} implies that
$H_s\pa{\paa{\la},\paa{\delta}}$ is a symmetric function of the
parameters $\la$ and of the parameters $\delta$. Moreover, $\text{e}^{2\pa{m-1}\la_j}\,\,
H_s\pa{\paa{\la},\paa{\delta}}$ is a polynomial in each of  the
$\text{e}^{2\la_j}$ variables of degree 2\pa{m-1}.
Finally, it is a matter of straightforward computations to check that $H_s$
satisfies the reduction properties:
\begin{multline}
 H_s\mid_{\la_1=\pm\pa{\delta_1-ia_1\zeta}}\pa{\paa{\la_i}_{i=1}^m;\paa{\delta_k}_{k=1}^{m}}=
 \pm H_s\pa{\paa{\la_i}_{i=2}^m;\paa{\delta_k}_{k=2}^{m}}  \\
\times \s{2\pa{\delta_1-i a_1\zeta}}\s{\delta_1 - \xi_-}
\pl_{j=2}^{m}\sd{\la_j,\delta_1+ia_1\zeta}\sd{\delta_1-2ia_1\zeta,\delta_k}.
\end{multline}
These are the reduction properties of
$\mc{Z}_{m}\pa{\paa{\la},\paa{\delta}}$, the partition function of
the six-vertex model with reflecting ends \cite{Tsu98}. Supplementing this result
with  the equality of the two functions at $m=1$, we obtain that
$H_s\pa{\paa{\la};\paa{\delta}}$ is
$s$-independent and equal to $\mc{Z}_m\pa{\paa{\la},\paa{\delta}}$.
Hence, we have the following result:
\begin{prop}[Boundary-type resummation]\label{prop-boundary-type-resum}
The generating function of the spin correlation function $\moy{ \mc{Q}_m\pa{\kappa}}$ can be expressed as
\begin{equation}
 \moy{ \mc{Q}_m\pa{\kappa}}= \sul_{s=0}^m \kappa^s F_s
\end{equation}
with\begin{multline}
F_s=\f{1}{s!\pa{m-s}!2^m}\Int{\mc{C}}{} \dd^m \la\
\f{\det{m}{\widetilde{\Psi}\pa{\la_i,\delta_j}}}{\pl_{i<j}\sd{\delta_i,\delta_j}}
\,\,
\pl_{j=1}^{m}\f{\s{\xi_-+\delta_j}}{\s{\la_j,\xi_- -ia_j\zeta}}
  \\
 \times
 \Theta_s\pa{\paa{\la}} \,\,  \mc{Z}_m\pa{\paa{\la},\paa{\delta}} \;\; .
\end{multline}
Here
\beq
\Theta_s\pa{\paa{\la}}=
\pl_{j>k}\f{\sd{\la_j,\la_k}}
{\sd{\la_{jk},i\ov{a}_{jk}\zeta}\sd{\ov{\la}_{jk},i\ov{a}_{jk}\zeta}} \;\; ,
\enq
and
\begin{multline}
\mc{Z}_m\pa{\paa{\la},\paa{\delta}}=
\frac{
\pl_{j,k=1}^m \big[\sd{\la_j,\delta_k+i\tf{\zeta}{2}}\,
\sd{\la_j,\delta_k-i\tf{\zeta}{2}}\big]}
{\pl_{i<j}\big[\sd{\la_i,\la_j}\,\sd{\delta_j,\delta_i}\big]}\\
\times \pl_{j=1}^{m}\big[\sinh 2\la_j\, \s{\delta_j-\xi_-}\big]\ \det{m}{\frac{1}{\sd{\la_j,\delta_k+i\tf{\zeta}{2}}
\sd{\la_j,\delta_k-i\tf{\zeta}{2}}}}.
\end{multline}
\end{prop}

Let us finally mention that the sum over $s$ in \eqref{sommesurpermuattion Fs} can be
rewritten as a single integral over auxiliary $z$ variables:
\beqa
\moy{\mc{Q}_m\pa{\kappa}} &=& \f{1}{m!\, 2^m} \Int{\mc{C}}{}
 \dd^m \la \hspace{-3mm}\oint\limits_{\bigcup\limits_{\eps=\pm} \Gamma_+\pa{\eps i \f{\zeta}{2}} }^{}\hspace{-3mm}  \f{\dd^m
z}{\pa{2i\pi}^m}\
\f{\mc{Z}_m\pa{\paa{\la},\paa{\delta}}\ \det{m}{\widetilde{\Psi}\pa{\la_i,\delta_j}}}
{\pl_{j>k}^{m}\big[\sd{\la_j,\la_k}
\sd{\la_{jk},i\zeta}\sd{\ov{\la}_{jk},i\zeta}\big]} \nonumber\\
 && \qquad  \times \pl_{k>l}^{m}
\f{\sd{\la_{kl},z_{kl}}\sd{\ov{\la}_{kl},z_{kl}}}
{\sd{\delta_l,\delta_k}} \;
\pl_{p=1}^{m} \f{ \varphi \pa{z_p} \s{\xi_-+\xi_p}}
{\sd{\la_p,\xi_-+z_p}},
\label{resommation compete Qkappa}
 \eeqa
where
\beq
\varphi\pa{z}= \f{\sinh 2z\,\,
\kappa^{-i\tf{\pa{z+i\tf{\zeta}{2}}}{\zeta}}}
{\sd{z,i\tf{\zeta}{2}}}.
\enq
\noindent This boundary-type resummation yields an integrand not only symmetric in $\paa{\la}$
but also invariant under a reversal of any integration variable $\la$. These properties allow to compute completely the so called emptiness formation probability at $\Delta=\tf{1}{2}$ and for vanishing boundary magnetic fields. It also allows to obtain the leading asymptotics of this quantity at the free fermion point \cite{Koz07}.

\section{The free fermion point}
\label{the free fermion point}
\subsection{Local magnetization at distance $m$}

The first (and the most important) application of the re-summation methods given above is the magnetization profile.
This one-point function $$\moy{ \sg_m^z}=1-2D_m
\partial_{\kappa} \moy{\mc{Q}_m\pa{\kappa}}\mid_{\kappa=1},$$  can be computed at the free fermion point by using the two different types of re-summations for the generation function. We give here both derivations.

\subsubsection{First method}
In the free fermion point,
 the  $\e{n}^{\e{th}}$ term of the series  \eqref{Q kappa original series} behaves as
$\pa{\kappa-1}^n$. Thus, after taking the $\kappa$ derivative and
sending $\kappa$ to 1, only the $n=1$ term survives. At $\zeta=\tf{\pi}{2}$, we have
\beqa
\langle\mc{Q}_m\pa{\kappa}\rangle&=& \sul_{n=0}^{m}
\f{\pa{\kappa-1}^n}{\pa{2\pi}^{2n}\pa{n!}^2}
\Int{\mc{C}_{D}}{} \dd^n\la
\oint\limits_{\Gamma_+\pa{\paa{\xi}}}\!\! \dd^n z\
\pl_{a=1}^{m}
\pl_{b=1}^{n}
\f{\tanh\pa{\la_b-\xi_a}}{\tanh\pa{z_b-\xi_a}}  \nonumber \\
&& \qquad \times \pl_{j=1}^{n} \paa{\f{\s{z_j+\xi_- +i \tf{\pi}{4}}}
{\s{\la_j+\xi_- +i \tf{\pi}{4}}}  \sinh 2\la_j } \nonumber \\
&& \qquad \times \det{n}{\f{1}{\s{\la_j-z_k}}}
\det{n}{\f{1}{\sd{\la_j,z_k}}}.
\eeqa
Thus \footnote{We do not give $\moy{\sg_1^z}$ as it corresponds to
$\partial_{\kappa} \moy{\mc{Q}_1\pa{\kappa}}$ without taking the
lattice derivative.}, for $m \geq 2$,
\beq
\moy{\sg_m^{z}}= \f{\pa{-1}^m}{\pi} \Int{\mc{C}_{D}}{}
\dd \la \,\,
\f{\s{\la-\xi_--i\tf{\pi}{4}}}{\s{\la+\xi_-+i\tf{\pi}{4}}}
\f{\,\,\,\pac{\tanh\pa{\la+i\tf{\pi}{4}}}^{2(m-1)}}{\cosh^2\pa{\la+i\tf{\pi}{4}}}.
\enq

\noindent Computing, if it exists, the residue at $\check{\la}$ we get,
for $m\geq 2$,
\begin{multline}
\moy{\sg_m^{z}}=-2\Theta\pa{h_- - 1} \f{h_-^2-1}{h_-^{2m}}
 \\
+ \f{\pa{-1}^m}{\pi} \Int{\R}{} \dd \la \,\,
\f{h_- + i \tanh\pa{\la+i\tf{\pi}{4}}}{1+i
h_-\tanh\pa{\la+i\tf{\pi}{4}}} \; \f{\,\,
\pac{\tanh\pa{\la+i\tf{\pi}{4}}}^{2(m-1)}}{\cosh^2\pa{\la+i\tf{\pi}{4}}},
\end{multline}
where $\Theta\pa{x}$ is the Heaviside step function. The standard
$\Delta=0$  change of variables,
\beq
\text{e}^{ip}=-\tanh\pa{\la-\tf{i\pi}{4}} \,\, ,
\enq
yields
\beq
\moy{\sg_m^z}=-2\Theta\pa{h_- - 1} \f{h_-^2-1}{h_-^{2m}}
 +\f{\pa{-1}^m}{\pi} \Int{0}{\pi} \dd p \,\,
\text{e}^{-2i\pa{m-1}p} \,\,\f{\text{e}^{-ip}+ih_- }{\text{e}^{ip}-ih_- }. \label{resultat exact pour sigma z}
\enq
Thus $\moy{\sg_m^z}$ displays Friedel type oscillations induced by the boundary.
Moreover it  decays as $\tf{1}{m}$ when $m\tend +\infty$:
\beq
\moy{\sg_m^z}=\f{2\pa{-1}^m}{\pi m} \f{h_-}{h_-^2+1} +
\text{O}\paf{1}{m^2}, \qquad m
>>1.
\enq

Here we recover the results of
\cite{Bil00a}, since we have
$h_-=\sqrt{2} \a_-$ in Bilstein's notations. When $\abs{h_-} \tend
\infty$ we conclude from \eqref{resultat exact pour sigma z} that
the first site is totally decoupled from the others as
$\moy{\sg_{m}^z}_{m\geq 2}$ goes to its bulk average value 0.
Actually in this limit the model is in correspondence
with a Kondo model with a spin $\tf{1}{2}$ impurity \cite{SaT95}.
But in this
case the impurity is completely screened,
and the overall magnetization in zero.

We also
recover from \eqref{resultat exact pour sigma z}, just as expected
from the spin reversal symmetry,  that $\moy{\sg_{m}^z}=0$ when the
boundary field vanishes. Actually this observation holds for
all $\Delta$ as inferred from the structure of the
monodromy matrix \eqref{Monodromy matrix Boundary} on the first site. When one sets
$\xi_1=\tf{\eta}{2}$ and $\xi_-=0$ then it acts as a diagonal matrix
on the first site, a sign of the claimed decoupling.

\subsubsection{Second method}
Starting from the  re-summation formula \eqref{Fs type Bulk} of Proposition~\ref{prop-bulk-type-resum}, we implement the simplification
due to $\zeta=\tf{\pi}{2}$. If we perform the change of variables
\beq
\ex{ip}=-\tanh\pa{\la-i\tf{\pi}{4}}
\enq
\noindent in \eqref{Fs type Bulk} at $\zeta=\tf{\pi}{2}$ then we arrive at
\begin{align}
F_s&=\f{\pa{2i}^{\f{m\pa{m-1}}{2}}}{s!\pa{m-s}!}\
  \pl_{j=1}^s\;
  \Int{\mc{C}_D}{} \f{dp_j}{2\pi}
  \pac{\f{\ex{i p_j}+\ex{-i p_j}}{\ex{i p_j}-ih_-}}
  \pl_{j=s+1}^m\;
  \Int{\ov{\mc{C}}_A}{} \f{dp_j}{2\pi}
  \pac{\f{\ex{i p_j}+\ex{-i p_j}}{\ex{-i p_j}-ih_-}}
   \nonumber\\
&\times
\pl_{k=1}^s\pl_{j=s+1}^m \big[ \pa{\ex{-i p_k}-\ex{i p_j}}\pa{\sin p_j +\sin p_k}\big]
\pl_{\substack{j,k=1 \\ j>k }}^{s}\big[ \pa{\ex{-i p_j}- \ex{-i p_k}}\pa{\sin p_j- \sin p_k}\big]
\nonumber\\
& \times
\pl_{\substack{j,k=s+1 \\ j>k}}^{m}\big[ \pa{\ex{i p_j}- \ex{i p_k}}\pa{\sin p_k - \sin p_j}\big]  .
\end{align}
The contours of integration are
\begin{alignat}{4}
&\ov{\mc{C}}_A=\intoo{0}{\pi},\quad & & h_- \geq -1,\qquad
&  &  \ov{\mc{C}}_A=\intoo{0}{\pi} \cup  \Gamma_-\pa{\ex{-ip}=-ih_-},\quad & & h_-<-1, \\
&\mc{C}_A=\intoo{-\pi}{0},\quad & & h_- \geq -1,\qquad
&  &  \mc{C}_A=\intoo{-\pi}{0} \cup  \Gamma_+\pa{\ex{ip}=-ih_-},\quad & & h_-<-1, \\
&\mc{C}_D=\intoo{0}{\pi},\quad &  & h_- \leq 1,\qquad
&  &  \mc{C}_D=\intoo{0}{\pi} \cup  \Gamma_+\pa{\ex{ip}=-ih_-},\quad & & h_- >1.
\end{alignat}

\noindent Once we introduce the function
\beq
\theta_{\kappa} \pa{p} = \left\{ \ba{cc }
                                \kappa & p \in \mc{C}_D, \\
                                    1 & p \in \mc{C}_A,
                            \ea \right.
\enq
\noindent  we can re-sum the terms $F_s$ into a single $m$-fold integral for $\moy{Q_m\pa{\kappa}}$:
\beqa
\moy{\mc{Q}_m\pa{\kappa}} & = &\f{\pa{2i}^{\f{m\pa{m-1}}{2}}}{m!}
\Int{\mc{C}_A \cup \, \mc{C}_D}{}\pl_{j=1}^m \pac{
 \f{\dd p_j}{2\pi}\,\theta_\kappa \pa{p_j}\,\f{\ex{i p_j}+\ex{-i p_j}}{\ex{i p_j}-ih_-} }
\nonumber\\
&&\hspace{1cm} \times
\pl_{ j>k\ge 1}^{m} \pa{\ex{-i p_j}- \ex{-i p_k}}\pa{\sin p_j- \sin p_k}.
\label{Q kappa fermions libres}
\eeqa
One can then express the generating function as a single determinant
\beqa
\moy{\mc{Q}_m \pa{\kappa}}&=&\det{m}{U\pa{\kappa}},\\
U_{j k}\pa{\kappa}&=& \f{1}{2\pi} \Int{\mc{C}_A \cup \,\mc{C}_D}{} \dd p \,\, \theta_\kappa\pa{p} \, \ex{-i p (j-1)}
\f{\ex{i p k}-(-1)^k \ex{-i p k}}{\ex{i p}-ih_-}.
\eeqa

\noindent To simplify this result we add to each row  of $U(\kappa)$ the next one multiplied by $ih_-$:
\begin{alignat*}{2}
\moy{\mc{Q}_m\pa{\kappa}}&=\mathrm{det}_m \widetilde{U}^{(m)}(\kappa),&&\\
\widetilde{U}^{(m)}_{j k}(\kappa)&= \f{1}{2\pi}\Int{\mc{C}_A \cup \,\mc{C}_D}{}
\dd p\,\,\theta_\kappa\pa{p}\,
\pa{\ex{i p (k-j)}-(-1)^k \ex{-i p (j+k)}},\quad&j<m&,\\
\widetilde{U}^{(m)}_{m k}(\kappa)&= U_{m k}(\kappa).&&
\end{alignat*}
It is easy to see  that $Q_m(1)=1$. Computing the first derivative of the generating function
one recovers the result already obtained from the series \eqref{resultat exact pour sigma z}:

\begin{align}
\moy{\sg_m^z} &= \f{(-1)^m}{\pi}\Int{\mc{C}_D}{} \dd p\,\,   \ex{-2i p (m-1)}\, \f{  \ex{-i p }+ih_-}{\ex{i p}-ih_-}.\nonumber\\
&=-2 \f{h_-^2-1}{h_-^{2m}} \Theta\pa{h_- - 1}
+\f{(-1)^m}{\pi} \Int{0}{\pi} \dd p\,\, \ex{-2i p (m-1)}\, \f{  \ex{-i p }+ih_-}{\ex{i p}-ih_-}.
\end{align}

\subsection{Local density of energy}

The local density of energy is another interesting quantity \cite{LafSCA06} that one can  evaluate for the XX0 chain:

\beq
E_m=\moy{\sigma^x_{m}\sigma^x_{m+1}+\sigma^y_{m}\sigma^y_{m+1}}\; .
\enq

\noindent Starting from \eqref{Es type Bulk} and  using the same technique as in the previous sub-section one easily obtains the following representation
for the density of energy:
\beq
E_m=-2i\Int{\mc{C}_D}{}\f{\dd p}{2\pi} \, \Int{\mc{C}_A}{}\f{\dd q}{2\pi} \,\pa{\ex{i(p+q)}+1}
\det{m+1}{V(p,q)}\; .
\enq
\noindent The entries of $V\pa{p,q}$ read
\begin{align}
&V_{jk}(p,q)=\f{1}{2\pi}\Int{\mc{C}_A \cup \,\mc{C}_D}{}
\hspace{-3mm} \dd p'\left(e^{i p' (k-j)}-(-1)^k e^{-i p' (j+k)}\right)=\delta_{j k},\qquad j<m-1,\nonumber\\
&V_{m-1 k}(p,q)=\f{1}{2\pi}\Int{\mc{C}_A \cup \,\mc{C}_D}{}
\hspace{-3mm}\dd p' \ex{-i p' (m-2)}
 \f{\ex{i p' k}-(-1)^k \ex{-i p' k}}{\ex{i p'}-ih_-} = \paf{h_-}{i}^{k-m+1} \hspace{-6mm}\Theta\pa{k- m+1},\nonumber\\
&V_{m k}(p,q)=\ex{-i p m} \f{\ex{i p k}-(-1)^k \ex{-i p k}}{\ex{i p}-ih_-},\nonumber\\
&V_{m+1 k}(p,q)=\ex{-i q m} \f{\ex{i q k}-(-1)^k \ex{-i q k}}{\ex{i q}-ih_-}.
\end{align}

\noindent Finally,
the integrals over $\mc{C}_A$ can be represented as
\beq
\Int{\mc{C}_A}{}=\Int{\mc{C}_A \cup \,\mc{C}_D}{}-\Int{\mc{C}_D}{} \; .
\enq
Accordingly, $E_m$ reduces to a sum of two $3 \times 3$ determinants:
\beqa
E_m&=&-2i \left| \ba{ccc}
                        1 & ih_-& -h_-^2\\
                        F(m-1, m-1)& F(m-1,m)&F(m-1,m+1)\\
                            0    & 1         &ih_-
                \ea \right|   \hspace{2cm} \nonumber\\
&& \hspace{2.5cm} -2i   \left| \ba{ccc}
                        1 & ih_-& -h_-^2\\
                        F(m, m-1)& F(m,m)&F(m,m+1)\\
                            0    &  0    &  1
                \ea\right| \; ,
\eeqa
\noindent where
\beq
F\pa{j,k}= \f{1}{2\pi}\Int{\mc{C}_D}{} \dd p \,\,\ex{-i p j} \,\,\f{\ex{i p k}-(-1)^k \ex{-i p k}}{\ex{i p}-ih_-}\; .
\enq

\noindent The computation of these determinants yields
\beq
E_m=-2i\pa{F(m,m)-F(m-1,m+1)}   -2h_-\pa{F(m,m-1)- F(m-1,m) } \; ,
\enq
\noindent or, more explicitly,

\beq
E_m=-\f{4}{\pi} + \f{2i}{\pi}\,
(-1)^m \, \Int{\mc{C}_D}{} \dd p \,\,\ex{-i p (2m-1)} \,\,\f{\ex{-i p }+ih_-}{\ex{i p}-ih_-} \; .
\enq

The constant term reproduces the bulk result. The influence of the boundary appears in
the oscillating term. In the
 $m\tend \infty$ limit the local density of energy behaves as

\beq
E_m=-\f{4}{\pi} + \f{2}{\pi m}\,(-1)^m\,\f{1-h_-^2}{1+h_-^2} + \e{O}\paf{1}{m^2}.
\enq

\subsubsection{Two-point function  $\moy{\sg_{m+1}^+\sg_{1}^-}$}

The preceding method can be successfully applied to compute other types
of two-point functions (like boundary-bulk two point functions), here we give the example of $\moy{\sg_{m+1}^+\sg_{1}^-}$.

In the free fermion point, after the usual change of variables and some straightforward but tedious calculations, one obtains from \eqref{Gs type Bulk}  a simple determinant formula for this object.
\begin{alignat}{2}
\moy{\sg_{m+1}^+\sg_{1}^-}=&-i \, \mathrm{det}_{m+1} V^{+-},&&\\
V^{+-}_{jk}=&\frac 1{2\pi i}\left(\int\limits_0^\pi-\int\limits_\pi^{2\pi}\right) d p\left(e^{i p (k-j-1)}-(-1)^k e^{-i p (j+1+k)}\right),\quad&j\le m-1&,\nonumber\\
V^{+-}_{m k}=&\frac 1{2\pi} \Int{\mc{C}_D}{} \dd p\, e^{-i p m} \frac {e^{i p k}-(-1)^k e^{-i p k}}{e^{i p}-ih_-},\quad&&\nonumber\\
V^{+-}_{m+1 k}=&\frac 1{2\pi}\int\limits_ {\mathcal{C}_A} d p\, e^{-i p m} \frac {e^{i p k}-(-1)^k e^{-i p k}}{e^{i p}-ih_-}&&
\end{alignat}
Computing the integrals in the first $m-1$ rows and using the fact that sum of the last two rows is $\delta_{k,m+1}$ we reduce this representation to a determinant of a $m\times m $ matrix
\begin{alignat}{2}
\moy{\sg_{m+1}^+\sg_{1}^-}=&\frac{i(-1)^m}{2 \pi^m} \, \mathrm{det}_{m} \tilde{V}^{+-},&&\\
 \tilde{V}^{+-}_{jk}=&(1+(-1)^{j-k})\frac {(j+1)(1-(-1)^{k})+k(1+(-1)^{k})}{(j+1)^2-k^2},\quad&j\le m-1&,\nonumber\\
 \tilde{V}^{+-}_{m k}=& \Int{\mc{C}_D}{} \dd p\, e^{-i p m} \frac {e^{i p k}-(-1)^k e^{-i p k}}{e^{i p}-ih_-},&&
 \label{determspsm}
\end{alignat}


This determinant can be computed for any value of $m$. However the result is quite different for $m$ odd or even. The details of the computation are given in Appendix B, here we give only the final result for the two point function
%
\begin{align}
\moy{\sg_{2a}^+\sg_{1}^-}=&-i\frac{2^{2a-1}}{\pi^2}\left(\pl_{j=1}^{2a-1}
\frac{\Gamma(j)}{\Gamma(j+\frac 12)}\right)^2\frac{\Gamma(a-\frac 12)\Gamma^3(a+\frac 12)}
{\Gamma(a)\Gamma(a+1)}\nonumber\\
&\times \Int{\mc{C}_D}{} \dd p\, P_{2a-1}\left(\vphantom{\prod}\cos p\right) \frac {e^{-i p (2a-1)}}{e^{i p}-ih_-}
\label{result_odd}\\
\moy{\sg_{2a+1}^+\sg_{1}^-}=&-\frac {2^{2a}}{\pi^2}\left(\pl_{j=1}^{2a}
\frac{\Gamma(j)}{\Gamma(j+\frac 12)}\right)^2\frac{\Gamma(a+\frac 32)\Gamma^3(a+\frac 12)}
{\Gamma(a)\Gamma(a+1)}\nonumber\\
&\times\int\limits_0^\pi d q\, \cos q \Int{\mc{C}_D}{} \dd p\, P_{2a}\left(\vphantom{\prod}\cos(q- p)\right) \frac {e^{-2i p a}}{e^{i p}-ih_-},
\label{result_even}
\end{align}
where $P_m(x)$ are  Legendre polynomials.

Asymptotic analysis of these expression yields the following leading behavior of $\moy{\sg_{m+1}^+\sg_1^-}$
\begin{align}
\moy{\sg_{m+1}^+\sg_{1}^-}=&(-1)^m A(h_-)\, m^{-\frac 34} \left(1+O(\frac 1{\sqrt{m}})\right)\\
\label{result_asym}
A(h_-)=&\sqrt{\frac 2{\pi(1+h_-^2) }} \exp\left\{
\frac14\int_0^\infty\frac{dt}t\left[e^{-4t}-
\frac1{\cosh^2t}\right]\right\}.
\end{align}


\section{Conclusion}

In this article we have obtained different types of physical correlation functions of the open $XXZ$ chain from re-summations of the multiple integrals
derived in \cite{KitKMNST07} for the elementary blocks.  At the free-fermion point, we were able to use these representations to derive explicit results such as the formula for the density
of energy profiles, a quantity arising in the study of quantum entanglement in spin chains 
\cite{LafSCA06}.

Just as in the bulk case, the question concerning the asymptotic behavior of the correlation functions outside of the free-fermion point  naturally arises. The problem is of the same order
of difficulty as in the bulk model. Indeed, the multiple integrals differ from their bulk counterparts only by factors due to the
$\mathbb{Z}_2$ symmetry $\la \tend -\la$ and the presence of boundary fields.

One could also wonder if it would be possible to tell something about the dynamical or temperature correlation functions. It seems that this
generalization is highly non-trivial.

Finally, we would like to stress that our expressions also simplify at other particular points such as $\Delta=\tf{1}{2}$. For instance, when $\Delta=\tf{1}{2}$, one can already compute completely the so-called emptiness formation probability when $h_-=0$ \cite{Koz07}.

\appendix

\section*{Appendices}
\renewcommand{\theprop}{\Alph{subsection}.\arabic{prop}}
\renewcommand{\thecor}{\Alph{subsection}.\arabic{cor}}
\renewcommand{\theequation}{\Alph{subsection}.\arabic{equation}}
\renewcommand{\thesubsection}{\Alph{subsection}}



\subsection{Asymptotic of the two-point function   $\moy{\sg_{m+1}^+\sg_{1}^-}$}
\setcounter{equation}{0}

In the last section we obtained a determinant representation \eqref{determspsm} for the  two-point function   $\moy{\sg_{m+1}^+\sg_{1}^-}$.
This determinant can be computed for any value of $m$. However the results are quite different for $m$ odd or even.

If $m$ is odd: $m=2a-1$, the result can be written in the following form
\begin{align}
 \mathrm{det}_{m} \tilde{V}^{+-}=& 2^{m+1}\pi^{m-3} \left(\pl_{j=1}^m
\frac{\Gamma(j)}{\Gamma(j+\frac 12)}\right)^2\frac{\Gamma(a-\frac 12)\Gamma^3(a+\frac 12)}
{\Gamma(a)\Gamma(a+1)}\nonumber\\
\times&\sul_{b=1}^{a}\frac{\Gamma(a-b+\frac 12)\Gamma(a+b-\frac 12)}
{\Gamma(a-b+1)\Gamma(a+b)}\nonumber\\
\times&\int\limits_0^\pi d p\, e^{-i p(2 a-1)} \frac {e^{i p (2b-1)}+ e^{-i p (2b-1)}}{e^{i p}-ih_-}
\label{det_m_odd}
\end{align}
If $m$ is even:  $m=2a$, the result is quite similar but there is a very important difference:
 \begin{align}
 \mathrm{det}_{m} \tilde{V}^{+-}=& 2^{m+1}\pi^{m-3} \left(\pl_{j=1}^m
\frac{\Gamma(j)}{\Gamma(j+\frac 12)}\right)^2\frac{\Gamma(a+\frac 32)\Gamma^3(a+\frac 12)}
{\Gamma(a)\Gamma(a+1)} \nonumber\\
\times& \sul_{b=1}^{a}\frac b{(b+\frac 12)(b-\frac 12)}
\frac{\Gamma(a-b+\frac 12)\Gamma(a+b+\frac 12)}
{\Gamma(a-b+1)\Gamma(a+b+1)}\nonumber\\
\times&\int\limits_0^\pi d p\, e^{-2i p a} \frac {e^{2i p b}- e^{-2i p b}}{e^{i p}-ih_-}
\label{det_m_even}
\end{align}

Asymptotic analysis of the prefactors in (\ref{det_m_odd}) and (\ref{det_m_even}) is rather simple, namely:
\begin{align}
\left(\pl_{j=1}^{2a-1}
\frac{\Gamma(j)}{\Gamma(j+\frac 12)}\right)^2&\frac{\Gamma(a-\frac 12)\Gamma^3(a+\frac 12)}
{\Gamma(a)\Gamma(a+1)}\nonumber\\
=&\frac {\sqrt{2}\pi} {2^m}m^{-\frac 14}\exp\left\{
\frac14\int_0^\infty\frac{dt}t\left[e^{-4t}-
\frac1{\cosh^2t}\right]\right\}\left(1+O(\frac 1m)\right)\\
\left(\pl_{j=1}^{2a}
\frac{\Gamma(j)}{\Gamma(j+\frac 12)}\right)^2&\frac{\Gamma(a+\frac 32)\Gamma^3(a+\frac 12)}
{\Gamma(a)\Gamma(a+1)} \nonumber\\
=&\frac  {\sqrt{2}\pi} {2^{m+1}}m^{\frac 34}\exp\left\{
\frac14\int_0^\infty\frac{dt}t\left[e^{-4t}-
\frac1{\cosh^2t}\right]\right\}\left(1+O(\frac 1m)\right)
\end{align}
For $m$ odd the sum in (\ref{det_m_odd}) can be rewritten 
 as follows:
\begin{align}
\sul_{b=1}^{a}\frac{\Gamma(a-b+\frac 12)\Gamma(a+b-\frac 12)}
{\Gamma(a-b+1)\Gamma(a+b)} &\left(e^{-2i p( a-b)}+e^{-2i p( a+b-1)}\right) \nonumber\\
=&\sul_{l=0}^{2a-1}\frac{\Gamma(l+\frac 12)\Gamma(2a-l-\frac 12)}
{\Gamma(l+1)\Gamma(2a-l)} e^{-2i pl},
\end{align}
and can be represented in terms of the Legendre polynomials $P_m(\cos p)$
\begin{align}
\sul_{l=0}^{2a-1}\frac{\Gamma(l+\frac 12)\Gamma(2a-l-\frac 12)}
{\Gamma(l+1)\Gamma(2a-l)} e^{-2i pl}=&\frac{\Gamma(2a-\frac 12)\Gamma(\frac 12)}
{\Gamma(2a)}\, \vphantom{F}_2 F_1(\frac 12,1-2a;\frac 32-2a; e^{-2i p})\nonumber\\
=&\pi e^{-ipm} P_m(\cos p)
\end{align}
Using Laplace asymptotic formula,
\begin{equation}
 P_m(\cos p)= \left( \frac {2}{\pi m\sin p}\right)^{\frac 12}
\cos\left[p\left(m+\frac 12\right) -\frac \pi 4\right]+O\left(\frac 1{m^{\frac 32}}\right),
\end{equation}
 for  the remaining integral we obtain the following leading term
\begin{equation}
\int\limits_0^\pi d p\, P_m(\cos p) \frac {e^{-i p m}}{e^{i p}-ih_-}=-i{\sqrt{\frac {\pi }{m(1+h_-^2) }}} \left(1+O(\frac 1{\sqrt{m}})\right)
\end{equation}
Assembling all the contributions we obtain the following leading term for the two-point function (for $m $ odd):
\begin{equation}
\moy{\sg_{m+1}^+\sg_{1}^-}=(-1)^m \sqrt{\frac {2}{\pi(1+h_-^2 )}} \exp\left\{
\frac14\int_0^\infty\!\frac{dt}t\left[e^{-4t}-
\frac1{\cosh^2t}\right]\right\}m^{-\frac 34}\! \left(\!1+O(\frac 1{\sqrt{m}})\!\right)
\label{as_result_odd}
\end{equation}

The same result holds for $m$ even, but the derivation is a little bit more tricky. The sum in (\ref{det_m_even}) can be once again rewritten in a more simple way
\begin{align}
 \sul_{b=1}^{a}&\frac b{(b+\frac 12)(b-\frac 12)}
\frac{\Gamma(a-b+\frac 12)(h_-)\Gamma(a+b+\frac 12)}
{\Gamma(a-b+1)\Gamma(a+b+1)}\int\limits_0^\pi d p\, e^{-2i p a} \frac {e^{2i p b}- e^{-2i p b}}{e^{i p}-ih_-}\nonumber\\
=&  \frac 12 \sul_{b=-a}^{a}\left(\frac 1{b+\frac 12}+\frac 1{b-\frac 12}\right)
\frac{\Gamma(a-b+\frac 12)\Gamma(a+b+\frac 12)}
{\Gamma(a-b+1)\Gamma(a+b+1)}\int\limits_0^\pi d p\, \frac {e^{2i p( b-a)}}{e^{i p}-ih_-}\nonumber\\
=&  i \sul_{b=-a}^{a}\frac{\Gamma(a-b+\frac 12)\Gamma(a+b+\frac 12)}
{\Gamma(a-b+1)\Gamma(a+b+1)}\int\limits_0^\pi d q\, e^{-2i q b} \cos q\int\limits_0^\pi d p\, \frac {e^{2i p( b-a)}}{e^{i p}-ih_-}\nonumber\\
=&  i \pi\int\limits_0^\pi d q\, \cos q\int\limits_0^\pi d p\,P_m(\cos(q-p)) \frac { e^{-imp} }{e^{i p}-ih_-}
\end{align}
where we introduced an additional integral to be able to express the result once again in terms of the Legendre polynomials. Asymptotic analysis of these integrals gives
\begin{equation}
 i \pi\int\limits_0^\pi d q\, \cos q\int\limits_0^\pi d p\,P_m(\cos(q-p)) \frac { e^{-imp} }{e^{i p}-ih_-}=-\left(\frac {\pi}{m}\right)^\frac 32\frac {2i}{\sqrt{1+h_-^2 }} \left(1+O(\frac 1{\sqrt{m}})\right),
\end{equation}
and it leads once again to the same leading term (\ref{as_result_odd}) for the two-point function.

\section*{Acknowledgments}

J.M. M., N. S. and V. T. are supported by CNRS.
N. K., K.K. K., J.M. M. and V. T. are supported  by the ANR programm GIMP ANR-05-BLAN-0029-01.
N. K., G. N. and V. T. are supported by the ANR programm MIB-05 JC05-52749.
N. S. is supported by the French-Russian Exchange Program, the
Program of RAS Mathematical Methods of the Nonlinear Dynamics, RFBR-05-01-00498, Scientific Schools 672.2006.1.
N. K., G. N. and N. S. would like to thank the Theoretical Physics group of the Laboratory of Physics at ENS Lyon for hospitality, which makes this collaboration possible.

\bibliographystyle{/home/vero/LPM/tex/TeX/style-biblio/h-physrev}


\end{document}